\documentclass[journal]{IEEEtran}
\usepackage{cite}  
\usepackage{graphicx}
\usepackage{algorithm}  
\usepackage{algpseudocode}  
\usepackage{latexsym}
\usepackage{amsmath,amssymb,amsfonts}  
\usepackage{bm}  
\usepackage{xcolor}
\usepackage[acronym,shortcuts]{glossaries}
\usepackage[font=footnotesize]{subcaption}
\usepackage[font=footnotesize]{caption}
\newacronym{UE}{UE}{user equipment}
\newacronym{BS}{BS}{base station}
\newacronym{XL-MIMO}{XL-MIMO}{extremely large-scale multiple-input-multiple-output}
\newacronym{MIMO}{MIMO}{multiple-input-multiple-output}
\newacronym{LoS}{LoS}{line-of-sight}
\newacronym{NLoS}{NLoS}{non-line-of-sight}
\newacronym{mmWave}{mmWave}{milimeter-wave}
\newacronym{cmWave}{cmWave}{centimeter-wave}
\newacronym{sub-THz}{sub-THz}{sub-terahertz}
\newacronym{CSI}{CSI}{channel state information}

\newacronym{ULA}{ULA}{uniform linear array}
\newacronym{QAM}{QAM}{quadrature amplitude modulation}
\newacronym{AoA}{AoA}{angles of arrival}
\newacronym{SOMP}{SOMP}{simultaneous orthogonal matching pursuit}
\newacronym{P-SOMP}{P-SOMP}{polar-domain simultaneous orthogonal matching pursuit}
\newacronym{OMP}{OMP}{orthogonal matching pursuit}
\newacronym{SIDCO}{SIDCO}{successive iterative decorrelation by convex optimization}
\newacronym{QPSK}{QPSK}{quadrature phase shift keying}
\newacronym{NMSE}{NMSE}{normalized mean-squared error}
\newacronym{BER}{BER}{bit error rate}
\newacronym{SNR}{SNR}{signal-to-noise ratio}
\newacronym{LS}{LS}{least square}
\newacronym{DFT}{DFT}{discrete Fourier transform}
\newacronym{SBL}{SBL}{sparse Bayesian learning}
\newacronym{EM}{EM}{expectation maximization}
\newacronym{ELBO}{ELBO}{evidence lower bound}
\newacronym{KL}{KL}{Kullback-Leibler}
\newacronym{MRC}{MRC}{maximum ratio combining}
\newacronym{LMMSE}{LMMSE}{linear minimum mean square error}
\newacronym{MIL}{MIL}{matrix inversion lemma}
\newacronym{FLOPs}{FLOPs}{floating point operations}
\newacronym{QC-SIDCO}{QC-SIDCO}{quadratic complex successive iterative decorrelation by convex optimization}
\newacronym{CoSaMP}{CoSaMP}{compressive sampling matching pursuit}

\newacronym{MDL}{MDL}{minimum description length}
\newacronym{AIC}{AIC}{Akaike information criterion}
\newacronym{BIC}{BIC}{Bayesian information criterion}

\newacronym{AoD}{AoD}{angles of depature}
\newacronym{THz}{THz}{sub-terahertz}
\newacronym{AWGN}{AWGN}{additive white Gaussian noise}
\newacronym{RF}{RF}{radio-frequency}
\newacronym{AoSA}{AoSA}{array-of-subarrays}
\newacronym{NF}{NF}{near-field}
\newacronym{FF}{FF}{far-field}
\newacronym{EP}{EP}{expectation propagation}
\newacronym{P-SIGW}{P-SIGW}{polar-domain simultaneous iterative gridless weighted}
\newacronym{MUSIC}{MUSIC}{multiple signal classification}

\newacronym{Soft-IC}{Soft-IC}{soft interference cancellation}
\newacronym{JCDE}{JCDE}{joint channel and data estimation}
\newacronym{BIP}{BIP}{bilinear inference problem}
\newacronym{BiGAMP}{BiGAMP}{bilinear generalized approximate message passing}
\newacronym{BP}{BP}{belief propagation}
\newacronym{CLT}{CLT}{central limit theorem}
\newacronym{BiGaBP}{BiGaBP}{bilinear Gaussian belief propagation}
\newacronym{BG}{BG}{Bernoulli Gaussian}

\begin{document}

\title{Joint Channel and Data Estimation for Multiuser Extremely Large-Scale MIMO Systems}

\author{Kabuto Arai,~\IEEEmembership{Graduate Student Member, IEEE},  Koji Ishibashi,~\IEEEmembership{Senior Member, IEEE},\\ Hiroki Iimori,~\IEEEmembership{Member, IEEE}, Paulo Valente Klaine, and Szabolcs Malomsoky
\thanks{K. Arai and K. Ishibashi are with the Advanced Wireless and Communication Research Center (AWCC), The University of Electro-Communications, Tokyo 182-8285, Japan (e-mail: k.arai@awcc.uec.ac.jp, koji@ieee.org)}
\thanks{H. Iimori, P. V. Klaine, and S. Malomsoky are with Ericsson Research, Ericsson Japan K.K., (e-mail: \{hiroki.iimori, paulo.valente.klaine, szabolcs.malomsoky\}@ericsson.com)}
~\nocite{2024Arai_NearCE}
}

%
\markboth{Journal of \LaTeX\ Class Files,~Vol.~14, No.~8, August~2021}%
{Shell \MakeLowercase{\textit{et al.}}: Sample article using IEEEtran.cls for IEEE Journals}


\maketitle

\begin{abstract}
    This paper proposes a \ac{JCDE} algorithm for uplink multiuser \ac{XL-MIMO} systems.
    The initial channel estimation is formulated as a sparse reconstruction problem based on the angle and distance sparsity under the near-field propagation condition.
    This problem is solved using non-orthogonal pilots through an efficient low complexity two-stage compressed sensing algorithm.
    Furthermore, the initial channel estimates are refined by employing a \ac{JCDE} framework driven by both non-orthogonal pilots and estimated data.
    The \ac{JCDE} problem is solved by sequential \ac{EP} algorithms, where the channel and data are alternately updated in an iterative manner.
    In the channel estimation phase, integrating Bayesian inference with a model-based deterministic approach provides precise estimations to effectively exploit the near-field characteristics in the beam-domain.
    In the data estimation phase, a \acf{LMMSE}-based filter is designed at each sub-array to address the correlation due to energy leakage in the beam-domain arising from the near-field effects.
    Numerical simulations reveal that the proposed initial channel estimation and \ac{JCDE} algorithm outperforms the state-of-the-art approaches in terms of channel estimation, data detection, and computational complexity.
\end{abstract}

\glsresetall


\IEEEpeerreviewmaketitle

\glsresetall

\section{Introduction}
    \Ac{XL-MIMO} has emerged as a promising technology, enabling sharp directive beamforming and extensive spatial multiplexing~\cite{2020Carvalho_XL_MIMO,2024Wang_XL_MIMO_tutorial}. 
    However, the significant increase in antenna aperture leads to an expansion of the Rayleigh distance \cite{2023Cui_NF_mag,2023Liu_NF_tutorial}, defined as the border between the near-field and far-field regions. 
    Thus, the near-field effects in \ac{XL-MIMO} systems may not be negligible in some practically-relevant circumstances, such as in small area coverage with high carrier frequency~\cite{2022Cui_PSOMP}.
    
    Unlike the conventional far-field, the near-field channel depends not only on angles but also on distances from signal sources such as \acp{UE} and scatterers.
    Hence, conventional channel estimation methods such as \cite{2018Rodriguez_SOMP,2018Hu_CSCE_superresolution}, which exploit the beam-domain sparsity under the assumption of planar wavefront, experience performance degradation in the near-field due to energy leakage effects in the beam-domain.
    %
    To tackle this issue, the authors in \cite{2022Cui_PSOMP} have proposed a \ac{P-SOMP} algorithm, which leverages the angle and distance sparsity known as polar-sparsity arising from the near-field peculiar characteristics.
    In \ac{P-SOMP}, polar (angle-distance) grids are generated by spatially quantizing the polar-domain to utilize compressed sensing techniques. 
    However, in multiuser systems, \ac{P-SOMP} requires orthogonal pilots to separate multiple \acp{UE}. 
    As such, this approach results in non-negligible overhead as the number of \acp{UE} grows, especially in \ac{XL-MIMO} systems capable of spatially multiplexing many \acp{UE}.
    
    Considering these challenges, a near-field channel estimation algorithm, which works even with non-orthogonal pilots, has been proposed in \cite{2024Xie_URA_2DCoSaMP}.
    However, due to the non-orthogonality among pilots, inter-user interference still remains, so it is necessary to jointly estimate all UE channel components.
    As a result, this joint estimation significantly increases computational complexity because it requires UE-wise polar grids, which leads to a large grid size.
    Therefore, the authors in \cite{2024Xie_URA_2DCoSaMP} proposed a 2D-\ac{CoSaMP} algorithm, which is based on the \ac{CoSaMP} algorithm in the polar-UE 2D domain constructed by UE-wise polar grids.
    While the 2D-CoSaMP algorithm can mitigate computational complexity, its estimation performance is hindered by overfitting to noisy measurements, which results from the inverse operation on over-sampled estimates.
    Consequently, subsequent data detection suffers from severe deterioration, particularly with high-order modulation.

    One of the prospective solutions to obtain accurate channel estimate with non-orthogonal pilots is \ac{JCDE} \cite{2024Ito_BiGaBP, 2023Ito_AoA,2019Yan_bilinear, 2019Chen_bilinear_cluster, 2022Iimori_bilinear_Grant_free}, where not only pilot sequences but also estimated data symbols are utilized as pilot replicas.
    The \ac{JCDE} problem can be formulated as a \ac{BIP}.
    One of the prominent algorithms based on a Bayesian framework for \ac{BIP} is \ac{BiGAMP} \cite{2014Parker_BiGAMP_part1}.
    \ac{BiGAMP} is an extension of GAMP, originally designed for a high-dimensional generalized-linear problem by utilizing loopy \ac{BP} with \ac{CLT} and Taylor-series approximations based on large system limit to simplify the \ac{BP} update\footnote{
        Assuming a bilinear inference problem from the noisy measurement $\mathbf{Y} = \mathbf{H} \mathbf{X} + \mathbf{N}$ with objective matrices $\mathbf{H} \in \mathbb{C}^{N \times U}$ and $\mathbf{X} \in \mathbb{C}^{U \times K}$, and a noise matrix $\mathbf{N} \in \mathbb{C}^{N \times K}$, 
        the large system limit means that the dimensions $N$, $U$, and $K$ tend to infinity while the rates $N/U$ and $K/U$ remain constant~\cite{2014Parker_BiGAMP_part1}.
    }.
    Due to the heavy dependency on the large system assumption of \ac{BiGAMP}, the convergence performance deteriorates significantly
    when the system size is insufficient, the pilot length is short, or the prior distribution is misspecified~\cite{2024Ito_BiGaBP}.
    To address these issues, the authors in \cite{2024Ito_BiGaBP} have proposed \ac{BiGaBP}, which relaxes the \ac{BP} update rules of \ac{BiGAMP}, based on GaBP \cite{2003Kabashima_GaBP}, without heavily relying on the approximation under a large system limit assumption.
    This relaxation of the approximation leads to performance improvements while maintaining the same complexity order as \ac{BiGAMP}.

    Bilinear inference algorithms that exploit physical model structures, such as channel sparsity in the beam-domain, have been investigated in \cite{2019Yan_bilinear, 2019Chen_bilinear_cluster}. 
    In these papers, the channel sparsity is modeled using a \ac{BG} prior distribution because this prior is analytically tractable with a closed-form posterior. 
    However, the sparse structure cannot be exactly expressed by the analytically tractable prior, which leads to modeling errors.
    To tackle this issue, the authors of \cite{2023Ito_AoA} have integrated a model-based deterministic approach \cite{2018Fan_AR} into a Bayesian inference framework \cite{2024Ito_BiGaBP}, referred to as AoA-aided \ac{BiGaBP}. This deterministic approach rectifies the model mismatch caused by the use of the tractable prior.

    However, since this method assumes a far-field model, the model correction via the deterministic approach is insufficient in the near-field, where the channel in the beam-domain exhibits cluster sparsity instead of simple sparsity with a sharp peak, due to energy leakage in the beam-domain.
    Moreover, AoA-aided \ac{BiGaBP} relies on a \ac{MRC}-based detector, which cannot address the correlation caused by the energy leakage.
    While the computational complexity of the MRC-based detection is relatively low, the data denoising process in AoA-aided BiGaBP, which relies on prior information about modulation constellations, requires significant computational complexity that scales proportionally not only to the modulation order but also to the number of antennas due to its reliance on the BP update rules.
    Hence, a channel estimation with a deterministic approach tailored to near-field structure, along with a computationally efficient data detection to address the energy leakage, is essential for the JCDE in the near-field region.
    %
    Furthermore, an accurate initial channel estimation is crucial for JCDE algorithms because the performance of data and channel estimation heavily depends on the accuracy of the initial estimate, where an inaccurate initial estimate might lead to error propagation in the JCDE iterations.
    
    Within the context outlined above, we propose a \ac{JCDE} algorithm for multiuser XL-MIMO systems with non-orthogonal pilots. 
    Our contributions are summarized as follows.
    \begin{itemize}
        \item {\bf Initial channel estimation for JCDE}: 
        A novel initialization mechanism for the multiuser near-field channel estimation problem with pilot contamination due to non-orthogonal pilots is proposed, 
        enabling an accurate initial estimate that is then used in the subsequent \ac{JCDE} algorithm as the starting values.
        The proposed initial channel estimation algorithm consists of two stages to maintain low computational complexity.
        In the first stage, angle and distance parameters for all UEs are estimated from the polar grids using the \ac{SOMP} algorithm without pairing between each estimated path and the corresponding UE.
        Subsequently, the second stage involves the UE-path pairing using 2D-OMP \cite{2011Yong_2DOMP} with a reduced number of grids constructed on the angle and distance parameters derived from the first stage.
        Owing to the above two-stage procedure, our proposed initial channel estimation outperforms the existing state-of-the-art scheme \cite{2024Xie_URA_2DCoSaMP}, while maintaining comparable computational complexity.
        \item {\bf JCDE algorithm with model-based estimation}: 
        A novel bilinear \ac{JCDE} inference algorithm is proposed, which integrates a model-based deterministic estimation mechanism with a Bayesian inference to address the modeling errors in a prior, as in~\cite{2024Ito_BiGaBP}. 
        In contrast to the state-of-the-art AoA-aided \ac{BiGaBP} under the assumption of far-field, our algorithm estimates the channels as an aggregation of two distinct quantities: 1) a model-based estimate that captures the near-field channel structure and 2) its modeling error that captures how different the current estimate is from the true channel.
        The model-based estimate is alternately updated through a matching pursuit algorithm exploiting the near-field model structures, whereas the residual modeling errors and data symbols are jointly estimated by the \ac{EP} algorithm \cite{2001Minka_EP}, where an approximate posterior is calculated by minimizing the \ac{KL} divergence.
        To tackle the correlation caused by the energy leakage in the beam-domain while reducing the complexity, we introduce a novel posterior calculation design that enables the implementation of a sub-array-wise \ac{LMMSE}-based filter, allowing parallel computation of the matrix inversion with a smaller dimension than the array size.
        This design results in lower computational complexity compared to the state-of-the-art method, owing to the modification of an extrinsic value generation that does not rely on BP rules.
    \end{itemize}

    %
    \textit{Notation}:
    The notation $[\mathbf{A}]_{i,j}$ indicates the $(i,j)$ element of the matrix $\mathbf{A}$.
    $(\cdot)^*$, $(\cdot)^\mathrm{T}$, and $(\cdot)^\mathrm{H}$ indicate conjugate, transpose, and conjugate transpose, respectively.
    A block diagonal matrix consisting of matrices $\mathbf{A}_1, \ldots \mathbf{A}_N$ is represented as $\mathrm{blkdiag}(\mathbf{A}_1, \ldots, \mathbf{A}_N)$.
    For a random variable $x$ and a probabilistic density function $p(x)$, $\mathbb{E}_{p(x)}[x]$ indicate the expectation of $x$ over $p(x)$.
    For any function $f(\mathbf{z})$, $\int_{\mathbf{z} / z_{i}} f(\mathbf{z})$ denotes the integral of $f(\mathbf{z})$ with respect to $\mathbf{z}$ except for $z_i$.
    The operator $\otimes$ denotes the Kronecker product.
    For the index sets $\mathcal{I} = \{1,2,\ldots,I\}$ and $\mathcal{J} = \{1,2,\ldots,J\}$, $\mathcal{I} \times \mathcal{J}$ denotes the cartesian product of $\mathcal{I}$ and $\mathcal{J}$.
    $\mathcal{I} \setminus i$ represent the set $\{1,\ldots,i-1,i+1,\ldots,I\}$.
    The notation $x \in [x_1, x_2]$ denotes that $x$ belongs to the closed interval between $x_1$ and $x_2$.

\vspace{-1ex}
\section{System Model}

\begin{figure}[t!]
        \centering
        \includegraphics[scale = 0.75]{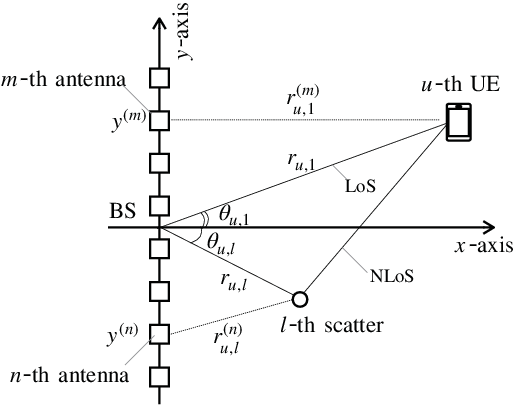}
        \caption{Near-field channel model. $r_{u,l}^{(n)}$ is the distance between the $n$-th antenna and the $u$-th UE ($l=1$) or scatterers ($l \neq 1$).}
        \label{fig:NF_model}
        \vspace{-3ex}
    \end{figure}

    We consider an uplink \ac{XL-MIMO} system, where a \ac{BS} has a \ac{ULA} with $N$-antennas, serving $U$ single antenna \acp{UE}.
    The \ac{ULA} is positioned along the $y$-axis, where~$y^{(n)} = (n-1)d - \frac{(N-1)d}{2},\ n=1,2,\ldots,N$ is the $n$-th antenna coordinate, and $d= \lambda / 2$ is antenna spacing with wavelength $\lambda$, as shown in Fig.\ref{fig:NF_model}.

    %
    \begin{figure}[t!]
        \centering
        \includegraphics[width = \linewidth]{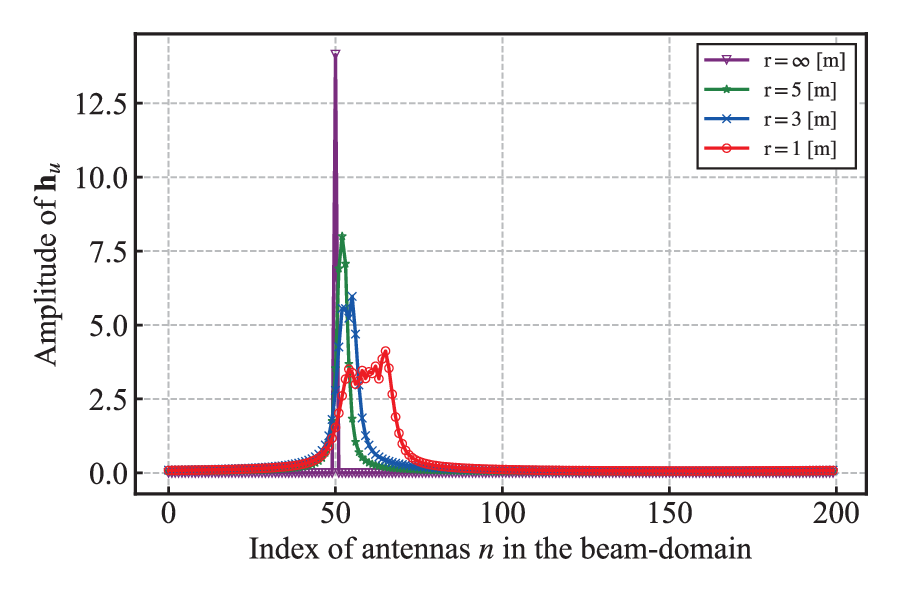}
        \caption{
        The amplitude of the channel vector in the beam-domain, $\mathbf{h}_u$ for $N=200, L=1, f_\mathrm{c}=100 \ \mathrm{GHz}$. 
        One UE is located $r$ meters away from the BS ($r=\infty$ corresponds to the far-field case).}
        \label{fig:NF_DFT}
        \vspace{-3ex}
    \end{figure}

    \vspace{-1ex}
    \subsection{Channel Model}
    \label{subsec:channel}

    The near-field channel in the spatial-domain between the \ac{BS} and the $u$-th UE is modeled as
    \begin{align}
        \label{eq:h_u}
        \mathbf{h}_u^\mathcal{S} = \sum_{l=1}^{L_u} \mathbf{a}(\theta_{u,l}, r_{u,l}) z_{u,l} = \mathbf{A}(\bm{\theta}_u, \mathbf{r}_u) \mathbf{z}_u,
    \end{align}
    where $\theta_{u,l} \in \mathbb{R}$,  and $z_{u,l} \in \mathbb{C}$ denote the \ac{AoA} and path gain of the $l$-th path and the $u$-th UE,
    and $L_u$ is the number of paths for the $u$-th UE~\cite{IimoriWCL20}.

    Without loss of generality, $l=1$ represents the \ac{LoS} component and $l \in \{2,\ldots L_u \}$ represents the \ac{NLoS} components.
    Let $L \triangleq \sum_{u=1}^U L_u$ denote the total number of paths including $U$-UEs. 
    Accordingly, $r_{u,1} \in \mathbb{R}$ denotes the distance between the \ac{BS} and the $u$-th \ac{UE}, and $r_{u,l} \in \mathbb{R},\ l \neq 1$ is the distance between the \ac{BS} and the $l$-th scatterer around the $u$-th \ac{UE}.
    Besides, $\mathbf{a}(\theta_{u,l}, r_{u,l}) \in \mathbb{C}^{N \times 1}$ denotes the array response vector defined as
    \begin{align}
        \label{eq:a_n}
        [\mathbf{a}(\theta_{u,l},r_{u,l})]_n
        = \exp \left[ 
            -j\frac{2 \pi}{\lambda} \left( r_{u,l}^{(n)} - r_{u,l} \right)
        \right],
    \end{align}
    where 
    $r_{u,l}^{(n)} = \sqrt{r_{u,l}^2 + y^{(n)2} -2 r_{u,l} y^{(n)} \sin \theta_{u,l}}$
    is the distance between the $n$-th antenna and the $u$-th UE ($l=1$) or scatterers ($l \neq 1$).

    For the $u$-th \ac{UE}, let us define the collections of \acp{AoA}, distances, and path gains as
    $\bm{\theta}_u \triangleq \{ \theta_{u,l} \}_{l=1}^{L_u} $, 
    $\mathbf{r}_u \triangleq \{ r_{u,l} \}_{l=1}^{L_u} $, and 
    $\mathbf{z}_u \triangleq \begin{bmatrix} z_{u,1}, \ldots, z_{u,L_u} \end{bmatrix}^\mathrm{T} \in \mathbb{C}^{L_u \times 1}$, respectively, and
    the corresponding array response matrix is defined as
    $\mathbf{A}(\bm{\theta}_u, \mathbf{r}_u) \triangleq \begin{bmatrix} \mathbf{a}(\theta_{u,1}, r_{u,1} ), \ldots, \mathbf{a}(\theta_{u,L_u}, r_{u,L_u}) \end{bmatrix} \in \mathbb{C}^{N \times L_u}$.
    Then, the channel matrix
    $\mathbf{H}^\mathcal{S} \triangleq \begin{bmatrix} \mathbf{h}_1^\mathcal{S}, \ldots , \mathbf{h}_U^\mathcal{S} \end{bmatrix} \in \mathbb{C}^{N \times U}$
    is written as 
    \begin{align}
        \label{eq:H}
        \mathbf{H}^\mathcal{S} = \mathbf{A}(\bm{\theta}, \mathbf{r}) \mathbf{Z},
    \end{align}
    where 
    $\mathbf{A}(\bm{\theta}, \mathbf{r}) \triangleq \begin{bmatrix} \mathbf{A}(\bm{\theta}_1, \mathbf{r}_1), \ldots, \mathbf{A}(\bm{\theta}_U, \mathbf{r}_U) \end{bmatrix} \in \mathbb{C}^{N \times L}$  
    is the array response matrix consisting of $U$-\acp{UE} with 
    AoAs, distances and path gains defined as
    $\bm{\theta} \triangleq \{\bm{\theta}_{u}\}_{u=1}^U$, 
    $\mathbf{r} \triangleq \{\mathbf{r}_{u}\}_{u=1}^U$, and 
    %
    $\mathbf{Z} \triangleq \mathrm{blkdiag} \begin{pmatrix} \mathbf{z}_1, \mathbf{z}_2, \ldots, \mathbf{z}_U \end{pmatrix} \in \mathbb{C}^{L \times U}$, respectively.

    The array response vector in the far-field region, \textit{i.e.,} when $r_{u,l} \rightarrow \infty $ is expressed as $[\mathbf{a}(\theta_{u,l},\infty)]_n = \exp 
    \left[  j\frac{2 \pi y^{(n)}}{\lambda} \sin \theta_{u,l} \right]$ from \eqref{eq:a_n}.
    As the far-field array response depends only on the angle, the far-field channel $\mathbf{h}_u^\mathcal{S}$ can be converted into a sparse beam-domain channel $\mathbf{h}_u = \mathbf{D}_N \mathbf{h}_u^\mathcal{S}$ with the \ac{DFT} matrix $\mathbf{D}_N \in \mathbb{C}^{N \times N}$.
    %
    In contrast, as the far-field approximation does not hold in the near-field region, the beam-domain near-field channel $\mathbf{h}_u$ exhibits not a simple sparse structure but rather a cluster sparse structure, which is caused by energy leakage due to a model mismatch between the DFT matrix $\mathbf{D}_N$ and the near-field array response $\mathbf{a}(\theta_{u,l},r_{u,l})$.        
    To illustrate the energy leakage effects, Fig.~\ref{fig:NF_DFT} depicts the amplitude of the beam-domain channel vector in the near-field and far-field regions.
    It can be seen that the far-field channel possesses a distinct sparse structure with a peaky spike. 
    On the other hand, the near-field channel exhibits a clustered sparsity with flatter peaks due to energy leakage.
    Hence, conventional channel estimation methods exploiting the beam-domain sparsity~\cite{2018Rodriguez_SOMP,2018Hu_CSCE_superresolution,2018Fan_AR} encounter significant performance degradation in the near-field.

    \subsection{Received Signal Model}
    \label{subsec:received}

    To estimate the near-field channel, the $u$-th UE transmits a pilot sequence $\mathbf{x}_{\mathrm{p},u} \in \mathbb{C}^{K_\mathrm{p} \times 1}$ and data symbol $\mathbf{x}_{\mathrm{d},u} \in \mathbb{C}^{K_\mathrm{d} \times 1}$ subsequently, where $K_\mathrm{p}$ and $K_\mathrm{d}$ are the length of pilots and data symbols\footnote{
    Since the pilots and data symbols are transmitted in distinct time symbols, unlike superimposed pilot transmission schemes~\cite{2004Lang_pilot_CE}, contamination between the pilots and data symbols does not occur.
    }.
    To reduce the pilot overhead for channel estimation, we employ non-orthogonal pilots that satisfy $K_\mathrm{p} < U$ and $\mathbf{x}_{\mathrm{p},u}^\mathrm{T} \mathbf{x}_{\mathrm{p},u^\prime}^\ast \neq 0, (u^\prime \neq u)$, which results in the pilot contamination between UEs unlike the orthogonal pilots.

    Each entry of $\mathbf{x}_{\mathrm{d},u}$ is randomly generated from a $Q$-\ac{QAM} constellation $ \mathcal{X} \triangleq \{\mathcal{X}_1, \ldots, \mathcal{X}_Q\}$ with average symbol energy $E_\mathrm{s}$.
    Then, the received pilot $\mathbf{Y}_\mathrm{p}^\mathcal{S} \in \mathbb{C}^{N \times K_\mathrm{p}}$ and data $\mathbf{Y}_\mathrm{d}^\mathcal{S} \in \mathbb{C}^{N \times K_\mathrm{d}}$ in the spatial-domain are given by 
    \begin{align}
        \label{eq:Y_p}
        \mathbf{Y}_\mathrm{p}^\mathcal{S} = \mathbf{H}^\mathcal{S} \mathbf{X}_\mathrm{p} + \mathbf{N}_\mathrm{p}^\mathcal{S},\ \ 
        \mathbf{Y}_\mathrm{d}^\mathcal{S} = \mathbf{H}^\mathcal{S} \mathbf{X}_\mathrm{d} + \mathbf{N}_\mathrm{d}^\mathcal{S}, 
    \end{align}
    where 
    $\mathbf{X}_\mathrm{p} \triangleq \begin{bmatrix} \mathbf{x}_{\mathrm{p},1}, \ldots, \mathbf{x}_{\mathrm{p},U} \end{bmatrix}^\mathrm{T} \in \mathbb{C}^{U \times K_\mathrm{p}}$ and 
    $\mathbf{X}_\mathrm{d} \triangleq \begin{bmatrix} \mathbf{x}_{\mathrm{d},1}, \ldots, \mathbf{x}_{\mathrm{d},U} \end{bmatrix}^\mathrm{T} \in \mathbb{C}^{U \times K_\mathrm{d}}$ 
    are the transmitted pilot matrix and data matrices.
    $\mathbf{N}_\mathrm{p} \in \mathbb{C}^{N \times K_\mathrm{p}}$ and 
    $\mathbf{N}_\mathrm{d} \in \mathbb{C}^{N \times K_\mathrm{d}}$ are the \ac{AWGN} matrices, whose entries are generated from $\mathcal{CN}(0, \sigma^2)$ with noise variance $\sigma^2$.
    By stacking the received pilot $\mathbf{Y}_\mathrm{p}^\mathcal{S}$ and data $\mathbf{Y}_\mathrm{d}^\mathcal{S}$, the effective received signal
    $\mathbf{Y}^\mathcal{S} \triangleq \begin{bmatrix} \mathbf{Y}_\mathrm{p}^\mathcal{S}, \mathbf{Y}_\mathrm{d}^\mathcal{S} \end{bmatrix} \in \mathbb{C}^{N \times K}$, with $K \triangleq K_\mathrm{p} + K_\mathrm{d}$, is formulated as
    \begin{align}
        \label{eq:Y_all}
        \mathbf{Y}^\mathcal{S} = \mathbf{H}^\mathcal{S} \mathbf{X} + \mathbf{N}^\mathcal{S},
    \end{align}
    with 
    $\mathbf{N}^\mathcal{S} \triangleq \begin{bmatrix} \mathbf{N}_\mathrm{p}^\mathcal{S}, \mathbf{N}_\mathrm{d}^\mathcal{S} \end{bmatrix} \in \mathbb{C}^{N \times K}$
    and 
    $\mathbf{X} \triangleq \begin{bmatrix} \mathbf{X}_\mathrm{p}, \mathbf{X}_\mathrm{d} \end{bmatrix} \in \mathbb{C}^{U \times K}$.
    For the sake of future convenience, let us define the pilot and data index set as 
    $ \mathcal{K} \triangleq \mathcal{K}_\mathrm{p} \cup \mathcal{K}_\mathrm{d}$ with 
    $ \mathcal{K}_\mathrm{p} \triangleq \{ 1,2, \ldots, K_\mathrm{p} \}$ and 
    $ \mathcal{K}_\mathrm{d} \triangleq \{ K_\mathrm{p} + 1, \ldots, K_\mathrm{p} + K_\mathrm{d} \}$.

    From \eqref{eq:H} and \eqref{eq:Y_p}, the received pilot $\mathbf{Y}_\mathrm{p}^\mathcal{S}$ is rewritten as
    \begin{align}
        \label{eq:Y_A}
        \mathbf{Y}^\mathcal{S}_\mathrm{p} &= \mathbf{A}(\bm{\theta}, \mathbf{r}) \mathbf{V} + \mathbf{N}_\mathrm{p}^\mathcal{S},
    \end{align}
    where 
    $\mathbf{V} \triangleq \mathbf{Z} \mathbf{X} = 
    \begin{bmatrix} \mathbf{x}_1 \mathbf{z}_1^\mathrm{T}, \mathbf{x}_2      
        \mathbf{z}_2^\mathrm{T}, \ldots, \mathbf{x}_U \mathbf{z}_U^\mathrm{T} 
    \end{bmatrix}^\mathrm{T} \in \mathbb{C}^{L \times K_\mathrm{p}}$ is the matrix composed of path gains and pilots.

\section{Overview of the Proposed Algorithm}
    \label{sec:overview}

\begin{figure}[t!]
        \centering
        \includegraphics[width=\linewidth]{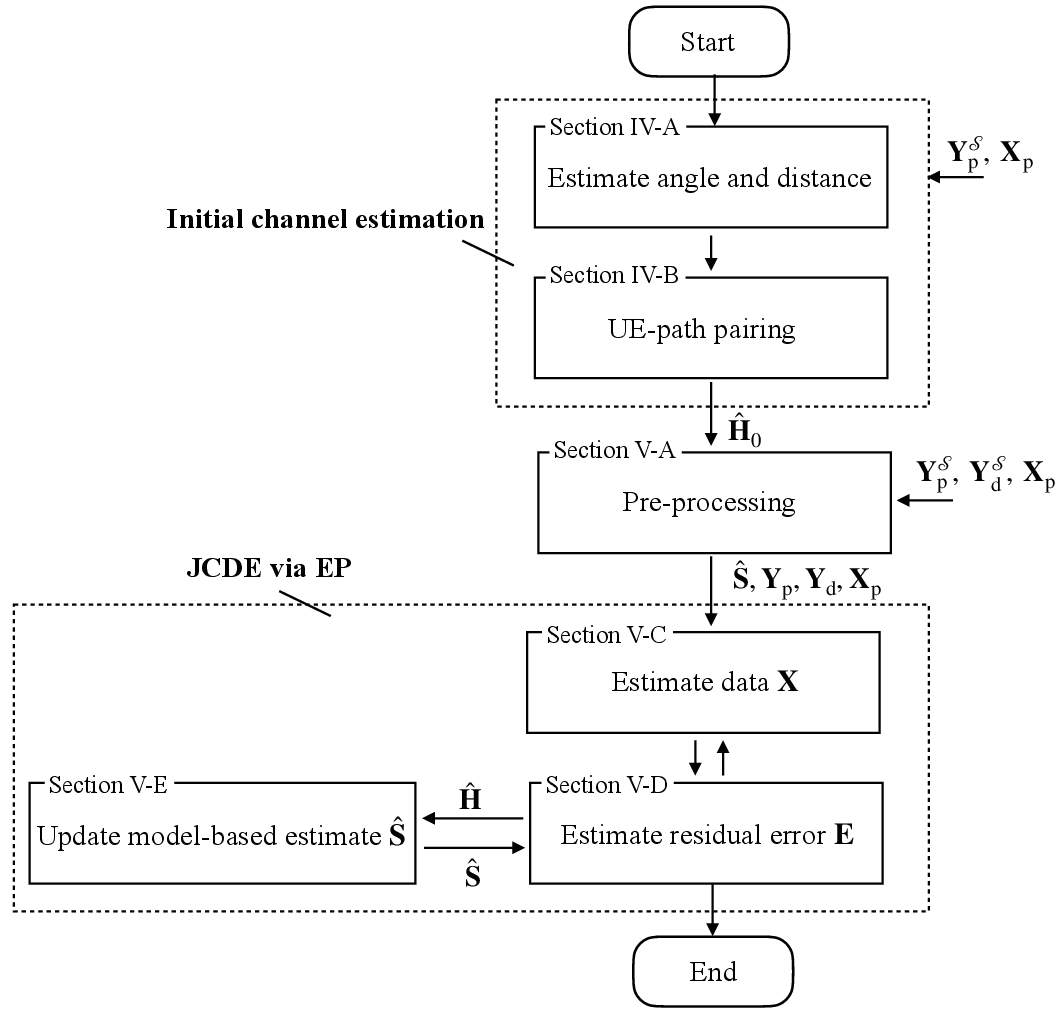}
        \caption{Overview of the proposed algorithm.}
        \label{fig:overview_algo}
    \end{figure}

    This section describes the overview of the proposed algorithm.
    The overall procedures of the proposed algorithm are illustrated in Fig.~\ref{fig:overview_algo}.
    As shown in the figure, the proposed algorithm mainly consists of two parts: the initial channel estimation part and subsequent \ac{JCDE} part.
    %
    The initial channel estimation part yields an accurate channel estimate to improve the convergence point of the subsequent \ac{JCDE} algorithm.
    This initial estimation is composed of two stages to reduce the computational complexity. 
    In the first stage, the angle and distance candidates from large-size polar grids are estimated by utilizing the \ac{SOMP} algorithm. 
    In the second stage, the pairing between the path candidates obtained in the first stage and corresponding UEs is performed via the 2D-OMP algorithm by using UE specific pilot sequences.
    The first and second stages for initial channel estimation are described in Section~\ref{subsec:First_stage} and Section~\ref{subsec:Second_stage}, respectively.
    
    In the subsequent \ac{JCDE} process, the channel and data are jointly estimated via the \ac{EP} algorithm with a deterministic model-based estimation approach, employing the initial channel estimate as the starting values.
    To exploit the near-field model structures, the beam-domain channel matrix $\mathbf{H}  \in \mathbb{C}^{N \times U}$ is decomposed into a model-based estimate $\hat{\mathbf{S}} \in \mathbb{C}^{N \times U}$ and residual channel error $\mathbf{E} \in \mathbb{C}^{N \times U}$.
    $\mathbf{E}$ and $\mathbf{X}$ are jointly estimated by the \ac{EP} algorithm, where the approximate joint posterior for $\mathbf{E}$ and $\mathbf{X}$ is calculated as described in Section~\ref{subsec:EP_x} and \ref{subsec:EP_e}.
    The model-based estimate $\hat{\mathbf{S}}$ is determined by the initial channel estimate and adaptively updated in the algorithm iterations to further improve estimation performance as described in Section~\ref{subsec:MB}.
    The performance of data and channel estimation in the JCDE algorithm heavily depends on the accuracy of the initial channel estimate.
    Therefore, creating an accurate initial channel estimate leads to a better convergence point in the JCDE algorithm, which is the main role of the initial channel estimation algorithm. 

\section{Proposed Initial Channel Estimation}
    \label{sec:initial_CE}

    \subsection{Angle and Distance Estimation}
    \label{subsec:First_stage}

    To leverage the near-field channel sparsity, the virtual channel representation in the polar-domain \cite{2018Rodriguez_SOMP} is utilized with polar grids.
    The polar grids are designed by spatially quantizing the angle and distance domain into $G_\theta G_r$ grid points as
    $\tilde{\bm{\theta}} \triangleq \{ \tilde{\theta}_{g_\theta} | g_\theta \in \{1,\ldots, G_\theta \} \}$ and 
    $\tilde{\mathbf{r}} \triangleq \left \{ \tilde{r}_{g_r, g_\theta} | g_r \in \{1, \ldots, G_r\},\ g_\theta \in \{1, \ldots, G_\theta\} \right \}$ 
    with $\tilde{\theta}_{g_\theta} \in [-\pi/2, \pi/2]$ and $\tilde{r}_{g_r, g_\theta} \in [0, \tilde{r}_\mathrm{max}]$\footnote{
        In this paper, we design the polar grids using the near-field sampling method proposed in~\cite{2024Xie_URA_2DCoSaMP}.
        Using this sampling method, the maximum distance grid is computed as $\tilde{r}_\mathrm{max} = 2N^2 \lambda \sqrt{\frac{0.001624}{1- \bar{\gamma}_\mathrm{coh}}}$ with a desired coherence value $\bar{\gamma}_\mathrm{coh}$, which is set to 0.6 in the numerical simulation in Section VI.
    }.
    Using the polar grids $\tilde{\bm{\theta}}$ and $\tilde{\bm{r}}$, the polar-domain dictionary (\textit{i.e.,} virtual array response matrix) is designed as
    %
    \begin{align}
        \label{eq:A_apx}
        \tilde{\mathbf{A}}( & \tilde{\bm{\theta}}, \tilde{\mathbf{r}}) = 
        \left[
            \mathbf{a}(\tilde{\theta}_1, \tilde{r}_{1, 1}), \ldots \mathbf{a}(\tilde{\theta}_1, \tilde{r}_{G_r, 1}), \ldots
        \right. \nonumber \\
        & \! \! \left.
            \ldots , 
            \mathbf{a}(\tilde{\theta}_{G_\theta}, \tilde{r}_{1, G_\theta}), \ldots ,\mathbf{a}(\tilde{\theta}_{G_\theta}, \tilde{r}_{G_r, G_\theta}) 
        \right] \in \mathbb{C}^{N \times G_\theta G_r}.
    \end{align}
    
    From \eqref{eq:Y_A} and \eqref{eq:A_apx}, the received pilot signal $\mathbf{Y}_\mathrm{p}^\mathcal{S}$ is given by
    \begin{align}
        \label{eq:Y_apx}
        \mathbf{Y}_\mathrm{p}^\mathcal{S} & \simeq \tilde{\mathbf{A}}( \tilde{\bm{\theta}}, \tilde{\mathbf{r}}) \tilde{\mathbf{V}} + \mathbf{N}^\mathcal{S}_\mathrm{p},
    \end{align}
    where 
    $\tilde{\mathbf{V}} \in \mathbb{C}^{G_\theta G_r \times K_\mathrm{p}}$ is the row sparse matrix such that the number of nonzero rows is only $L$ and other $G_\theta G_r - L$ rows are zero since the channel is composed of a total of $L$ paths defined as in \eqref{eq:h_u}, with a sufficiently large number of grids, \textit{i.e.,} $G_\theta G_r\gg L$.
    Equation \eqref{eq:Y_apx} exactly holds only if there is no quantization errors in polar grids.
    In actual environments, however, it approximately holds due to the presence of quantization errors.
    Therefore, to compensate the quantization errors, we overestimate the number of paths $(\hat{L} > L)$\footnote{
    This paper assumes that prior knowledge of the number of paths can be obtained from long-term statistics because the temporal variation in the number of paths is significantly slower than that of the channel. This assumption has been widely adopted in various studies~\cite{2022Cui_PSOMP, 2024Lei_Hybrid_CE}.
    }.
    To estimate $\hat{L}$ path candidates from $G_\theta G_r$ grids, the sparse reconstruction problem for $\tilde{\mathbf{V}}$ is formulated as
    \begin{align}
        \label{eq:min_ini_1}
        & \underset{\tilde{\mathbf{V}}}{\text{minimize}}
        \ \  \left \| 
            \mathbf{Y}_\mathrm{p}^\mathcal{S} - \tilde{\mathbf{A}}( \tilde{\bm{\theta}}, \tilde{\mathbf{r}}) \tilde{\mathbf{V}}
        \right \|_{\mathrm{F}}^2  \nonumber \\
        & \text{subject to}
        \ \  \| \tilde{\mathbf{V}} \|_{2,0} = \hat{L},
    \end{align}
    where 
    $ \| \tilde{\mathbf{V}} \|_{2,0}$ denotes the number of non-zero rows of $\tilde{\mathbf{V}}$.

    The problem in \eqref{eq:min_ini_1} can be approximately solved by a compressed sensing algorithm for multiple measurement vectors (MMV) problems, \textit{e.g.,} \ac{SOMP}~\cite{2018Rodriguez_SOMP},
    because $\tilde{\mathbf{V}}$ has common support across its columns (\textit{i.e.,} along the pilot dimension), meaning that the index set of nonzero elements is identical for all column vectors of $\tilde{\mathbf{V}}$.
    %
    The complexity of \ac{SOMP} at the $t$-th iteration in a naive implementation is $ \mathcal{O} (G_r G_\theta N K_\mathrm{p} + Nt + N t^2 +t^3)$.
    Its complexity can be further reduced by using the \ac{MIL} to $ \mathcal{O} (G_r G_\theta K_\mathrm{p} t + N t)$ \cite{1993pati_OMP_MIL, 2012Bob_OMP_Comparison}.
    %
    Solving the problem \eqref{eq:min_ini_1} yields the angle and distance candidates corresponding to the non-zero rows of $\tilde{\mathbf{V}}$, defined as 
    $\check{\bm{\theta}} \triangleq \{ \check{\theta}_l \}_{l=1}^{\hat{L}} $ and 
    $\check{\mathbf{r}} \triangleq \{ \check{r}_l \}_{l=1}^{\hat{L}} $.

    \subsection{UE-Path Pairing}
    \label{subsec:Second_stage}

    The path candidate set $\{\check{\theta}_l, \check{r}_l \}_{l=1}^{\hat{L}}$ obtained in the first stage does not specify the association of individual paths with each UE.
    To estimate individual channels for each UE, the second stage performs UE-path pairing, where the estimated path candidates are associated with each user using UE-specific non-orthogonal pilot sequences $\{\mathbf{x}_{\mathrm{p},u} \}_{u=1}^U$.
    %
    The usage of limited path candidates $\check{\bm{\theta}}$, $\check{\mathbf{r}}$, rather than large-size polar grids $\tilde{\bm{\theta}}$, $\tilde{\mathbf{r}}$ sampling the entire polar-domain, can lead to a complexity reduction.
    Using the path set $\{\check{\theta}_l, \check{r}_l \}_{l=1}^{\hat{L}}$, the polar-domain dictionary matrix is designed as
    \begin{align}
        \label{eq:A_candidate}
        \check{\mathbf{A}}(\check{\bm{\theta}}, \check{\mathbf{r}}) \! \triangleq  \! \begin{bmatrix} \mathbf{a}(\check{\theta}_1, \check{r}_1), \mathbf{a}(\check{\theta}_2, \check{r}_2), \ldots \mathbf{a}(\check{\theta}_{\hat{L}}, \check{r}_{\hat{L}}) \end{bmatrix} \in \mathbb{C}^{N \times \hat{L}}.
    \end{align}

    Reducing the size of the polar grids from $G_r G_\theta$ in \eqref{eq:A_apx} to $\hat{L} (\ll G_r G_\theta)$ in \eqref{eq:A_candidate} can effectively lower the complexity in the following compressed sensing algorithm.
    Then, the channel vector for the $u$-th UE can be approximated with the polar-domain dictionary $\check{\mathbf{A}}(\check{\bm{\theta}}, \check{\mathbf{r}})$ as
    \begin{align}
        \label{eq:h_u_apx}
        \mathbf{h}_u^\mathcal{S} = \mathbf{A}(\bm{\theta}_u, \mathbf{r}_u) \mathbf{z}_u \simeq \check{\mathbf{A}}(\check{\bm{\theta}}, \check{\mathbf{r}}) \check{\mathbf{z}}_u,
    \end{align}
    where $\check{\mathbf{z}}_u \in \mathbb{C}^{\hat{L} \times 1}$ is the virtual path gain vector.
    
    From \eqref{eq:h_u_apx}, the received pilot $\mathbf{Y}_\mathrm{p}^\mathcal{S}$ is approximated as
    \begin{align}
        \mathbf{Y}_\mathrm{p}^\mathcal{S}
        & \simeq \sum_{u=1}^{U} \check{\mathbf{A}}(\check{\bm{\theta}}, \check{\mathbf{r}}) \check{\mathbf{z}}_u \mathbf{x}_{\mathrm{p},u}^\mathrm{T} \!+\! \mathbf{N}_\mathrm{p} 
        \label{eq:Yp_apx_2}
        = \check{\mathbf{A}}(\check{\bm{\theta}}, \check{\mathbf{r}}) \check{\mathbf{Z}} \mathbf{X}_\mathrm{p} + \mathbf{N}_\mathrm{p}^\mathcal{S},
    \end{align}
    with $\check{\mathbf{Z}} \triangleq 
    \begin{bmatrix}
        \check{\mathbf{z}}_1, \check{\mathbf{z}}_2, \ldots,\   \check{\mathbf{z}}_U
    \end{bmatrix} \in \mathbb{C}^{\hat{L} \times U}$.

    The equation \eqref{eq:Yp_apx_2} can be transformed into a 1D linear equation as 
    $ \mathbf{y}_\mathrm{p}^\mathcal{S} \simeq \mathbf{\Phi}_\mathrm{p} \check{\mathbf{z}}$
    with 
    $\mathbf{y}_\mathrm{p}^\mathcal{S} \triangleq \mathrm{vec}(\mathbf{Y}^\mathcal{S}_\mathrm{p})$, 
    $\check{\mathbf{z}} \triangleq \mathrm{vec}(\check{\mathbf{Z}}) \in \mathbb{C}^{\hat{L}U \times 1}$ and 
    $ \mathbf{\Phi}_\mathrm{p} \triangleq \left ( \mathbf{X}_\mathrm{p}^\mathrm{T} \otimes \check{\mathbf{A}}(\check{\bm{\theta}}, \check{\mathbf{r}}) \right ) \in \mathbb{C}^{N K_\mathrm{p} \times \hat{L} U} $.
    Although the estimation for $\check{\mathbf{z}}$ from the vectorized observation $ \mathbf{y}_\mathrm{p}^\mathcal{S}$ can be simply addressed by various methods such as OMP \cite{1993pati_OMP_MIL}, this significantly increases the complexity due to the large-size dictionary $\mathbf{\Phi}_\mathrm{p}$.
    Hence, to circumvent the high computational burden, the 2D signal representation in \eqref{eq:Yp_apx_2} is directly addressed without the vectorized 1D representation.
    Then, the sparse reconstruction problem for $\check{\mathbf{Z}}$ in \eqref{eq:Yp_apx_2} is formulated as
    \begin{align}
        \label{eq:min_ini_2}
        & \underset{\check{\mathbf{Z}}}{\text{minimize}} \ \  
        \left \|  
            \mathbf{Y}_\mathrm{p}^\mathcal{S} - \check{\mathbf{A}}(\check{\bm{\theta}}, \check{\mathbf{r}}) \check{\mathbf{Z}} \mathbf{X}_\mathrm{p}
        \right \|_{\mathrm{2}}^2 \nonumber \\
        & \text{subject to}
        \ \  \left \| \check{\mathbf{z}}\right \|_{0} = \hat{L},
    \end{align}
    with $\check{\mathbf{z}} \triangleq \mathrm{vec}(\check{\mathbf{Z}}) \in \mathbb{C}^{\hat{L}U \times 1} $.

    The optimization problem \eqref{eq:min_ini_2} is solved via a two-dimensional compressed sensing algorithm. 
    The conventional method \cite{2024Xie_URA_2DCoSaMP} tackles this problem with the large-size polar dictionary $\tilde{\mathbf{A}}( \tilde{\bm{\theta}}, \tilde{\mathbf{r}})$ in \eqref{eq:A_apx} instead of $\check{\mathbf{A}}(\check{\bm{\theta}}, \check{\mathbf{r}})$ in \eqref{eq:A_candidate} via the 2D-\ac{CoSaMP} algorithm, which sacrifices estimation performance for complexity reduction compared to 2D-OMP \cite{2011Yong_2DOMP}.
    In contrast, our proposed method solves the optimization problem \eqref{eq:min_ini_2} via the 2D-OMP algorithm using the small-size polar-domain dictionary $\check{\mathbf{A}}(\check{\bm{\theta}}, \check{\mathbf{r}})$ constructed by the path candidates $\{\check{\theta}_l, \check{r}_l \}_{l=1}^{\hat{L}}$ in the first stage.
    As a result, the proposed method possesses the prominent capability to overcome the conventional approach \cite{2024Xie_URA_2DCoSaMP} while retaining comparable computational complexity.
    Detailed discussions regarding the complexity of the proposed algorithm are presented in Section \ref{sec:simulation}.

    Solving the problem \eqref{eq:min_ini_2} yields the estimated path gain vector $\hat{\mathbf{z}}_u \in \mathbb{C}^{\hat{L}_u \times 1}$, angle $\hat{\bm{\theta}}_u \in \mathbb{R}^{\hat{L}_u \times 1}$, and distance $\hat{\mathbf{r}}_u \in \mathbb{R}^{\hat{L}_u \times 1}$ corresponding to the non-zero elements of $\check{\mathbf{z}}_u \in \mathbb{C}^{\hat{L} \times 1}$, where $\hat{L}_u$ is the estimated number of paths for the $u$-th UE.
    Given the estimates, the initial channel estimate can be obtained as 
    \begin{align}
        \label{eq:H_est}
        \hat{\mathbf{H}}^{\mathcal{S}}_0 \!= \! \begin{bmatrix} \hat{\mathbf{h}}_1^\mathcal{S}, \ldots, \hat{\mathbf{h}}_U^\mathcal{S} \end{bmatrix}  \in \mathbb{C}^{N \times U},
        \text{with } 
        \hat{\mathbf{h}}_u^\mathcal{S} \! = \! \hat{\mathbf{A}}(\hat{\bm{\theta}}_u, \hat{\mathbf{r}}_u) \hat{\mathbf{z}}_u,
    \end{align}
    where $\hat{\mathbf{A}}(\hat{\bm{\theta}}_u, \hat{\mathbf{r}}_u) =
    \begin{bmatrix} 
        \mathbf{a}(\hat{\theta}_{u,1}, \hat{r}_{u,1}), \ldots, \mathbf{a}(\hat{\theta}_{u,\hat{L}_u}, \hat{r}_{u,\hat{L}_u})
    \end{bmatrix} \in \mathbb{C}^{N \times \hat{L}_u}$ is the estimated array response.
    The proposed initial channel estimation method is summarized in Algorithm~\ref{alg:initial}.

    \begin{algorithm}[t!]
\caption[]{Proposed channel estimation algorithm}
\label{alg:initial}
\hrulefill
\begin{algorithmic}[1]
    \vspace{-0.5ex}
    \Statex \textbf{Input:} 
        $\mathbf{Y}_\mathrm{p}^\mathcal{S},\ 
        \mathbf{X}_\mathrm{p},\  
        \hat{L}$
    \Statex \textbf{Output:} 
        $\hat{\mathbf{H}}_0^\mathcal{S}, \{\hat{\bm{\theta}}_u, \hat{\mathbf{r}}_u, \hat{\mathbf{z}}_u,\}_{u=1}^U$
    \vspace{-1.5ex}
    \Statex \hspace{-3ex} \hrulefill
    %
    \Statex 
        \textbf{// First Stage - Angle and distance estimation} 
    \State
        Calculate polar-domain dictionary $\tilde{\mathbf{A}}( \tilde{\bm{\theta}}, \tilde{\mathbf{r}})$ in \eqref{eq:A_apx}
    \State
        Estimate $\check{\bm{\theta}}, \check{\mathbf{r}}$ by solving \eqref{eq:min_ini_1} via \ac{SOMP} with \ac{MIL} \cite{1993pati_OMP_MIL}
    \Statex 
        \textbf{// Second Stage - UE-path pairing} 
    \State
        Calculate polar-domain dictionary $\check{\mathbf{A}}(\check{\bm{\theta}}, \check{\mathbf{r}})$ in \eqref{eq:A_candidate}
    \State
        Estimate $\{ \hat{\bm{\theta}}_u, \hat{\mathbf{r}}_u, \hat{\mathbf{z}}_u \}_{u=1}^U$ by solving \eqref{eq:min_ini_2} via 2D-OMP~\cite{2011Yong_2DOMP}
    \State 
        Estimate channel matrix $\hat{\mathbf{H}}_0^\mathcal{S}$ in \eqref{eq:H_est}
    \vspace{-0.5ex}
\end{algorithmic}
\end{algorithm}
\setlength{\textfloatsep}{5pt} 

\section{Proposed joint channel and data estimation}
    \label{sec:JCDE}

    Given the initial estimates obtained from Algorithm~\ref{alg:initial}, we aim to improve both the channel and data estimation performance while considering near-field properties.
    This section elaborates on the proposed JCDE algorithm using the initial channel estimate as the starting values. 

    \vspace{-2ex}
    \subsection{Pre-processing for Channel and Data Estimation}
    \label{subsec:preprocessing_data}

    \subsubsection{Pre-processing for Channel Estimation}

    To exploit the channel sparsity, the received signal and channel matrix in the spatial-domain are transformed in the beam-domain as 
    $\mathbf{Y} \triangleq \mathbf{D}_N \mathbf{Y}^\mathcal{S} \in \mathbb{C}^{N \times K}$ and 
    $\mathbf{H} \triangleq \mathbf{D}_N \mathbf{H}^\mathcal{S} \in \mathbb{C}^{N \times U}$.
    {
    Since the DFT matrix $\mathbf{D}_N$ is unitary, the transformation between the spatial and beam-domains can be efficiently performed via simple matrix multiplication unlike the polar-domain representation.
    Therefore, the matrix size remains unchanged after the transformation, resulting in no additional computational complexity in the subsequent iterative channel estimation and data detection processes.
    }
    However, as described in Section~\ref{subsec:channel}, the near-field channel in the beam-domain has a cluster sparse structure due to energy leakage. 
    Thus, to tackle this problem, 
    the channel matrix $\mathbf{H}$ is first considered as the aggregation of the model-based estimate $\hat{\mathbf{S}}\in \mathbb{C}^{N \times U}$ and the residual channel estimation error $\mathbf{E} \triangleq \mathbf{H} - \hat{\mathbf{S}}\in \mathbb{C}^{N \times U}$, resulting in 
    \begin{align}
        \label{eq:H_E}
        \mathbf{H} = \hat{\mathbf{S}} + \mathbf{E}.
    \end{align}
    
    An initial value for the model-based estimate $\hat{\mathbf{S}}$ is determined with the proposed initial channel estimate $\hat{\mathbf{H}}_0^\mathcal{S}$ in \eqref{eq:H_est} as $\hat{\mathbf{S}} = \mathbf{D}_N \hat{\mathbf{H}}_0^\mathcal{S}$, and it is adaptively updated based on the near-field model structure as described in Section~\ref{subsec:MB}.
    As the residual error $\mathbf{E}$ is defined by subtracting the current estimate $\hat{\mathbf{S}}$ from the beam-domain channel $\mathbf{H}$
    as in \eqref{eq:H_E}, this subtraction results in a sparser domain representation compared to the original beam-domain channel $\mathbf{H}$. The dominant path components are removed from $\mathbf{H}$ by $\hat{\mathbf{S}}$, facilitating the sparse matrix reconstruction by considering $\mathbf{E}$ (instead of $\mathbf{H}$) as the variable to be estimated by a Bayesian inference framework.

    \subsubsection{Pre-processing for Data Estimation}

    For low-complexity data estimation, the conventional methods based on the far-field assumption, such as \cite{2019Yan_bilinear,2023Ito_AoA,2019Chen_bilinear_cluster}, utilize \ac{MRC}-based detectors, which are ineffective in the near-field scenario because the near-field channel has cluster sparsity due to energy leakage, and the leaked energy is correlated in the beam-domain.
    Although \ac{LMMSE}-based detection methods such as \cite{2017Ma_OAMP, 2019Rangan_VAMP} are effective to deal with the correlation, these methods require matrix inversion with the size $N$, which is computationally expensive especially in XL-MIMO systems.
    To balance the computational complexity and detection performance, the array is virtually divided into multiple sub-arrays, and a sub-array-wise LMMSE-based detector is designed similarly to \cite{2020Wang_EP_subarray}.
    In contrast to \cite{2020Wang_EP_subarray}, which assumes perfect \ac{CSI}, the proposed method considers the channel estimation error while jointly estimating data and channel, exploiting the near-field model structures.

    Accordingly, the extra-large array with $N$ antennas are partitioned into $C$ sub-arrays, and the sub-array $c \in \mathcal{C} \triangleq \{1,2,\ldots,C \}$ has $N_c$ antennas satisfying $N = \sum_{c=1}^C N_c$.
    The received signals, residual channel errors, and model-based estimates can be also seen as
    $\mathbf{Y} =  \begin{bmatrix} \mathbf{Y}_1^\mathrm{T}, \mathbf{Y}_2^\mathrm{T}, \dots, \mathbf{Y}_C^\mathrm{T} \end{bmatrix}^\mathrm{T}$,
    $\mathbf{E} =  \begin{bmatrix} \mathbf{E}_1^\mathrm{T}, \mathbf{E}_2^\mathrm{T}, \dots, \mathbf{E}_C^\mathrm{T} \end{bmatrix}^\mathrm{T}$, and 
    $\hat{\mathbf{S}} =  \begin{bmatrix} \hat{\mathbf{S}}_1^\mathrm{T}, \hat{\mathbf{S}}_2^\mathrm{T}, \dots, \hat{\mathbf{S}}_C^\mathrm{T} \end{bmatrix}^\mathrm{T} $,
    with 
    $\mathbf{Y}_c \in \mathbb{C}^{N_c \times K}$, 
    $\mathbf{E}_c \in \mathbb{C}^{N_c \times U}$, and 
    $\hat{\mathbf{S}}_c \in \mathbb{C}^{N_c \times U}$.
    The received signals $\mathbf{Y}$ and $\mathbf{Y}_c$ can then be rewritten as
    \begin{align}
        \label{eq:Y_E}
        \mathbf{Y} = \mathbf{E} \mathbf{X} + \hat{\mathbf{S}} \mathbf{X} + \mathbf{N},
        \text{ with }
        \mathbf{Y}_c = \mathbf{E}_c \mathbf{X} + \hat{\mathbf{S}}_c \mathbf{X} + \mathbf{N}_c.
    \end{align}
    
    For convenience, let us define 
    $\mathcal{N} \triangleq \{1,2,\ldots,N\}$ as the antenna index set, and 
    $\mathcal{N}_c \triangleq \left \{ {n_{c(1)}}, n_{c(2)}, \ldots, n_{c(Nc)} \right \} \subset \mathcal{N}$ as the antenna index set at the $c$-th sub-array such that $\mathcal{N}_1 \cup \mathcal{N}_2 \cup \cdots \cup \mathcal{N}_C = \mathcal{N}$ and $\mathcal{N}_i \cap \mathcal{N}_j = \emptyset,\ i \neq j \in \mathcal{C}$.

    \subsection{Bayesian Inference Formulation}

    Based on the linear observation in \eqref{eq:Y_E} with the deterministic variable $\hat{\mathbf{S}}$ and random variables $\mathbf{X}$ and $\mathbf{E}$, 
    the likelihood function for $\mathbf{X}$ and $\mathbf{E}$ can be expressed as
    \begin{align}
        \label{eq:likelihood}
        p(\mathbf{Y}|\mathbf{E}, \mathbf{X}) 
        \!=\! \prod_{n \in \mathcal{N}} \prod_{k \in \mathcal{K}} p(y_{n,k} | \bar{\mathbf{x}}_k, \bar{\mathbf{e}}_n),
    \end{align}
    where
    $p(y_{n,k} | \bar{\mathbf{e}}_n, \bar{\mathbf{x}}_k) = \mathcal{CN} ( (\bar{\mathbf{e}}_n + \bar{\mathbf{s}}_n)^\mathrm{T} \bar{\mathbf{x}}_k,\ \sigma^2 )$
    with
    $\bar{\mathbf{e}}_{n} = [e_{n, 1}, \ldots, e_{n, U}]^\mathrm{T} \in \mathbb{C}^{U \times 1}$, 
    $\bar{\mathbf{s}}_{n} = [\hat{s}_{n, 1}, \ldots, \hat{s}_{n, U}]^\mathrm{T} \in \mathbb{C}^{U \times 1}$, and 
    $\bar{\mathbf{x}}_{k} = [x_{1, k}, \ldots, x_{U, k}]^\mathrm{T} \in \mathbb{C}^{U \times 1}$.

    Since each entry of $\mathbf{X}_\mathrm{d}$ is randomly selected from the \ac{QAM} constellation point set $ \mathcal{X}$, the prior $p(\mathbf{X})$ can be written as
    \begin{align}
        \label{eq:prior_x}
        p(\mathbf{X}) = \prod_{u \in \mathcal{U}} \prod_{k \in \mathcal{K}} p(x_{u,k}),
    \end{align}
    with
    $p(x_{u,k_d}) = \frac{1}{Q} \sum_{\mathcal{X}_i \in \mathcal{X}} \delta(x_{u,k_d} - \mathcal{X}_i),\ \forall k_d \in \mathcal{K}_d$ and
    $p(x_{u,k_p}) = \delta(x_{u,k_p} - [\mathbf{X}_\mathrm{p}]_{u,k_p}),\ \forall k_p \in \mathcal{K}_p$.

    Although many conventional methods such as \cite{2019Yan_bilinear,2019Chen_bilinear_cluster} design the i.i.d. sparse prior for the beam-domain channel as $p(\mathbf{H}) = \prod_{n \in \mathcal{N}} \prod_{u \in \mathcal{U}} p(h_{n,u})$ (\textit{e.g.,} \ac{BG} prior), this modeling causes the model mismatch due to energy leakage effects in the near-field region.
    Therefore, we design the sparse prior for the residual channel error $\mathbf{E}$ instead of $\mathbf{H}$ as 
    \begin{align}
        \label{eq:prior_e}
        p(\mathbf{E};\mathbf{\Theta}) = \prod_{n \in \mathcal{N}} \prod_{u \in \mathcal{U}}  p(e_{n,u};\mathbf{\Theta}),
    \end{align}
    where 
    $p(e_{n,u};\mathbf{\Theta}) = \mathcal{CN}(0,\ \sigma^e_{n,u})$ is Gaussian prior distribution with zero mean and variance $\sigma^e_{n,u}$, which is widely used for sparse representation in the \ac{SBL} algorithm 
    \cite{2006Bishop_PRML}, 
    where
    $\mathbf{\Theta} \triangleq \{ \sigma^e_{n,u} \}_{n \in \mathcal{N}, u \in \mathcal{U}}$ is the hyperparameter set to be optimized through the \ac{EM} algorithm \cite{2006Bishop_PRML} as described in Section~\ref{subsec:EM}.
    From the likelihood in \eqref{eq:likelihood} and priors in \eqref{eq:prior_x}, \eqref{eq:prior_e}, the posterior can be written as
    \begin{align}
        \label{eq:joint}
        p(\mathbf{E}, \mathbf{X} | \mathbf{Y};\mathbf{\Theta}) 
        = p(\mathbf{Y}|\mathbf{E}, \mathbf{X}) p(\mathbf{X}) p(\mathbf{E};\mathbf{\Theta}) / p(\mathbf{Y};\mathbf{\Theta}),
    \end{align}
    where
    $p(\mathbf{Y};\mathbf{\Theta}) = \int_{\mathbf{E},\mathbf{X}}
    p(\mathbf{Y}, \mathbf{E}, \mathbf{X};\mathbf{\Theta}) $
    is the marginal likelihood referred to as the evidence for parameter $\mathbf{\Theta}$.
    Our objective is to estimate $\mathbf{E}$, $\mathbf{X}$, and $\mathbf{\Theta}$ through the posterior and the evidence.
    
    The estimator for $\mathbf{\Theta}$ by the type-II maximum likelihood method \cite{2001Tipping_SBL} is given as 
    \begin{align}
        \label{eq:min_ML}
        \hat{\mathbf{\Theta}} = \underset{\mathbf{\Theta}}{\mathrm{argmax}}\ 
        p(\mathbf{Y};\mathbf{\Theta}).
    \end{align}
    
    However, the calculation of the evidence $p(\mathbf{Y};\mathbf{\Theta})$ is intractable due to the multidimensional integral for $\mathbf{X}$ and $\mathbf{E}$.
    Hence, we utilize the \ac{EM} algorithm, which maximizes the \ac{ELBO} in each iteration,
    instead of directly maximizing the evidence \cite{2006Bishop_PRML}.
    Given $\mathbf{\Theta}^{(t)}$ at the $t$-th iteration, $\mathbf{\Theta}^{(t+1)}$ at the $(t+1)$-th iteration can be obtained as the following E-step and M-step:
    \begin{align}
        & \text{E-step : } 
        \mathcal{F} ( \mathbf{\Theta}, \mathbf{\Theta}^{(t)} ) \! = \!
        \mathbb{E}_{p(\mathbf{E}, \mathbf{X} | \mathbf{Y} ; \mathbf{\Theta}^{(t)}) } \!
        \left [ \ln p(\mathbf{Y}, \mathbf{E}, \mathbf{X};\mathbf{\Theta})  \right] 
        + \mathsf{c}_0^{(t)} \!, \nonumber \\
        \label{eq:EM_M}
        &\text{M-step : } 
        \mathbf{\Theta}^{(t+1)} = \underset{\mathbf{\Theta}}{\mathrm{argmax}}\  \mathcal{F} ( \mathbf{\Theta}, \mathbf{\Theta}^{(t)} ),
    \end{align}
    where 
    $\mathcal{F} ( \mathbf{\Theta}, \mathbf{\Theta}^{(t)} )$ is the \ac{ELBO} with the constant value 
    $\mathsf{c}_0^{(t)} = \mathbb{E}_{p(\mathbf{E}, \mathbf{X} | \mathbf{Y} ; \mathbf{\Theta}^{(t)}) } 
    \left [ \ln p(\mathbf{E}, \mathbf{X} | \mathbf{Y} ;\mathbf{\Theta}^{(t)}) \right] $.
    
    Since E-step requires the calculation of a multidimensional integral that is computationally unreasonable, we approximate the posterior by $g^{(t)}(\mathbf{E}, \mathbf{X}| \mathbf{Y}) \simeq p(\mathbf{E}, \mathbf{X} | \mathbf{Y} ; \mathbf{\Theta}^{(t)})$, using the \ac{EP} algorithm.
    After the E-step, the maximization problem in \eqref{eq:EM_M} with the approximate posterior $g^{(t)}(\mathbf{E}, \mathbf{X}| \mathbf{Y})$ is solved, which is described in detail in Section~\ref{subsec:EM}.
    The EP procedure continues until it reaches the maximum number of iterations $T$.
    Finally, the last updated parameters at $t=T$ are used as the final estimates as 
    $\hat{\mathbf{\Theta}} \triangleq \mathbf{\Theta}^{(T)}$, 
    $\hat{\mathbf{E}} \triangleq \mathbb{E}_{g^{(T)}(\mathbf{E}, \mathbf{X}| \mathbf{Y})} [\mathbf{E}]$, and 
    $\hat{\mathbf{X}} \triangleq \mathbb{E}_{g^{(T)}(\mathbf{E}, \mathbf{X}| \mathbf{Y})} [\mathbf{X}] $.
    For notation simplicity, let us drop the iteration index $t$.
    The approximate posterior $g(\mathbf{E}, \mathbf{X} | \mathbf{Y})$ is derived by \ac{KL} minimization subject to a Gaussian distribution set $\mathbf{\Phi}$ as
    \begin{align}
        \label{eq:min_KL}
        \underset{g \in \mathbf{\Phi}}{\mathrm{minimize}}\  \mathrm{KL} \left(
        p(\mathbf{E}, \mathbf{X} | \mathbf{Y};\mathbf{\Theta}) \| 
        g(\mathbf{E}, \mathbf{X} | \mathbf{Y}) \right), 
    \end{align}
    where 
    the approximate posterior $g(\mathbf{E}, \mathbf{X} | \mathbf{Y})$ is designed as 
    \begin{align}
        \label{eq:g_apx}
        g(\mathbf{E}, \mathbf{X} | \mathbf{Y}) 
        &= Z_g^{-1} Q^x(\mathbf{X}) Q^e(\mathbf{E}) B^x(\mathbf{X}) B^e(\mathbf{E}),
    \end{align}
    where
    $Z_g = \int_{\mathbf{E}, \mathbf{X}} Q^x(\mathbf{X}) Q^e(\mathbf{E}) B^x(\mathbf{X}) B^e(\mathbf{E})$ is a normalizing constant, and
    $Q^x(\mathbf{X})$, $Q^e(\mathbf{E})$, $B^x(\mathbf{X})$, and $B^e(\mathbf{E})$ are the approximate factors such that 
    $Q^x(\mathbf{X}) Q^e(\mathbf{E}) \simeq p(\mathbf{Y} | \mathbf{E}, \mathbf{X})$,
    $B^x(\mathbf{X}) \simeq p(\mathbf{X})$, and
    $B^e(\mathbf{E}) \simeq p(\mathbf{E};\bm{\Theta})$ subject to Gaussian distribution set $\mathbf{\Phi}$.

    \setcounter{equation}{34}
    \begin{figure*}
        \begin{align}
            \label{eq:px_cond}
            \bar{p}^x_{c,u,k} (\mathbf{y}_{c,k} | x_{u,k})
            &\triangleq
            \bar{C}^x_{c,u,k}
            \int_{\mathbf{E}_c, \bar{\mathbf{x}}_k \setminus x_{u,k}}
            p(\mathbf{y}_{c,k} | \mathbf{E}_c, \bar{\mathbf{x}}_k) 
            \prod_{u^\prime \in \mathcal{U} \setminus u} v^x_{c,u^\prime,k}(x_{u^\prime,k})
            \prod_{u^\prime \in \mathcal{U}}
            \prod_{n_c \in \mathcal{N}_c} v^e_{n_c,u^\prime,k}(e_{n_c,u^\prime}), \\
            \label{eq:v_e_1}
            v^e_{n_c,u, k} (e_{n_c,u})
            & \triangleq 
            b_{n_c,u}^e(e_{n_c,u}) \prod_{k^\prime \in \mathcal{K} \setminus k} q^e_{n_c,u,k^\prime}(e_{n_c,u}) = 
            C^{v,e}_{n_c,u,k} 
            \exp \left ( - |e_{n_c,u} - \hat{e}^{v}_{n_c,u,k} |^2 / \xi^{v,e}_{n_c,u,k} \right ), \\
            \tilde{v}^e_{c,u,k} (\mathbf{e}_{c,u})
            & \triangleq \prod_{n_c \in \mathcal{N}_c} v^e_{n_c,u,k}(e_{n_c,u}) \propto 
            \exp \left\{- 
            (\mathbf{e}_{c,u} - \hat{\mathbf{e}}^v_{c,u,k})^\mathrm{H} 
            \mathbf{\Xi}^{e,v-1}_{c,u,k} 
            (\mathbf{e}_{c,u} - \hat{\mathbf{e}}^v_{c,u,k})
            \right\}.
            \label{eq:ve_exp}
        \end{align}
        \hrulefill
    \end{figure*}
    \setcounter{equation}{25}

    These approximate factors are designed as
    $Q^x(\mathbf{X}) = \prod_{c,u,k} q^x_{c,u,k}(x_{u,k})$, 
    $Q^e(\mathbf{E}) = \prod_{c} \prod_{n_c,u,k} q^e_{n_c,u,k}(e_{n_c,u})$,  
    $B^x(\mathbf{X}) = \prod_{u,k} b^x_{u,k}(x_{u,k})$, 
    $B^e(\mathbf{E}) = \prod_{c} \prod_{n_c, u} b^e_{n_c,u}(e_{n_c,u})$, 
    where $q^x_{c,u,k}(\cdot)$, $q^e_{n_c,u,k}(\cdot)$, $b^x_{u,k}(\cdot)$, and $b^e_{u,k}(\cdot)$ are the parameterized approximate functions defined as
    \begin{subequations}
    \begin{align}
        \label{eq:qx}
        q^x_{c,u,k}(x_{u,k}) & \triangleq \exp \left ( - |x_{u,k} - \hat{x}^{q}_{c,u,k} |^2 / \xi^{q,x}_{c,u,k} \right ), \\
        \label{eq:qe}
        q^e_{n_c,u,k}(e_{n_c,u}) &\triangleq \exp \left ( - |e_{n_c,u} - \hat{e}^q_{n_c,u,k} |^2 / \xi^{q,e}_{n_c,u,k} \right ), \\
        \label{eq:bx}
        b^x_{u,k}(x_{u,k}) &\triangleq \exp \left ( - |x_{u,k} - \hat{x}^{b}_{u,k} |^2 / \xi^{b,x}_{u,k} \right ), \\
        \label{eq:be}
        b^e_{n_c,u}(e_{n_c,u}) &\triangleq \exp \left ( - |e_{n_c,u} - \hat{e}^b_{n_c,u} |^2 / \xi^{b,e}_{n_c,u} \right ), 
    \end{align}
    \end{subequations}
    where 
    $\bm{\pi}^{q,x}_{c,u,k} \!\triangleq\! \begin{bmatrix} \hat{x}^{q}_{c,u,k}, \xi^{q,x}_{c,u,k} \end{bmatrix}^\mathrm{T}$,
    $\bm{\pi}^{q,e}_{n_c,u,k} \!\triangleq\! \begin{bmatrix} \hat{e}^{q}_{n_c,u,k}, \xi^{q,e}_{n_c,u,k}\end{bmatrix}^\mathrm{T}$, 
    $\bm{\pi}^{b,x}_{u,k} \!\triangleq\! \begin{bmatrix} \hat{x}^{b}_{u,k}, \xi^{b,x}_{u,k} \end{bmatrix}^\mathrm{T}$, and 
    $\bm{\pi}^{b,e}_{n_c,u} \!\triangleq\! \begin{bmatrix} \hat{e}^{b}_{n_c,u}, \xi^{b,e}_{n_c,u} \end{bmatrix}^\mathrm{T}$
    are unknown parameters to be optimized by \ac{KL} minimization.

    Since the approximate posterior $g(\mathbf{E}, \mathbf{X} |\mathbf{Y})$ in \eqref{eq:g_apx} is designed subject to Gaussian distribution set $\mathbf{\Phi}$, 
    the marginalized approximate posterior $g(x_{u,k} | \mathbf{Y}) $ and $g(e_{n_c, u} | \mathbf{Y}) $ can be expressed as
    $g(x_{u,k} | \mathbf{Y}) = \mathcal{CN}(\hat{x}_{u,k}, \xi^x_{u,k})$ and 
    $g(e_{n_c,u} | \mathbf{Y}) = \mathcal{CN}(\hat{e}_{n_c,u}, \xi^e_{n_c,u})$,
    where $\hat{x}_{u,k}$ and $\hat{e}_{n_c,u}$ are the posterior means, and $\xi^x_{u,k}$ and $\xi^e_{n_c,u}$ are the posterior variances.

    Let $\mathbf{\Pi} \triangleq \{\bm{\pi}^{q,x}_{c,u,k},\ \boldsymbol{\pi}^{b,x}_{u,k},\ \boldsymbol{\pi}^{q,e}_{n_c,u,k},\ \boldsymbol{\pi}^{b,e}_{n_c,u} \}_{ c \in \mathcal{C}, n_c \in \mathcal{N}, u \in \mathcal{U}, k \in \mathcal{K}}$
    denote an unknown parameter set.
    The optimal parameter set $\hat{\mathbf{\Pi}}$ is obtained by KL minimization in \eqref{eq:min_KL}.
    However, the objective function cannot be expressed in closed-form because 
    $\mathrm{KL} \left( p(\mathbf{E}, \mathbf{X} | \mathbf{Y};\mathbf{\Theta}) \| g(\mathbf{E}, \mathbf{X} | \mathbf{Y}) \right)$ 
    includes intractable integral with respect to the true posterior $p(\mathbf{E}, \mathbf{X} | \mathbf{Y};\mathbf{\Theta})$.
    To tackle this, we set the target distribution $\hat{p}(\mathbf{E}, \mathbf{X} | \mathbf{Y})$ instead of the true posterior into the KL divergence in \eqref{eq:min_KL}.
    The target distribution is designed by replacing a part of the true posterior with the approximate functions in \eqref{eq:qx}-\eqref{eq:be} as described in the following sections.
    For the sake of notation for the design of the target distribution, the approximate distribution
    $l^x_{c, k} (\mathbf{E}_{c}, \bar{\mathbf{x}}_k)  \simeq p(\mathbf{y}_{c,k} | \mathbf{E}_c, \bar{\mathbf{x}}_k)$ and
    $l^e_{n_c, k} (\bar{\mathbf{e}}_{n_c}, \bar{\mathbf{x}}_k) \simeq p(y_{n_c,k}| \bar{\mathbf{e}}_{n_c}, \bar{\mathbf{x}}_k )$
    are expressed using \eqref{eq:qx}-\eqref{eq:be} as
    \begin{subequations}
    \begin{align}
        \label{eq:likelihood_apx_x}
        &\hspace{-1.5ex}l^x_{c, k} (\mathbf{E}_{c}, \bar{\mathbf{x}}_k) \! \propto \! \prod_{u \in \mathcal{U}} q^x_{c,u,k}(x_{u,k}) \prod_{u \in \mathcal{U}} \prod_{n_c \in \mathcal{N}_c} \! q^e_{n_c,u,k}(e_{n_c,u}), \! \! \!  \\
        \label{eq:likelihood_apx_e}
        &\hspace{-1.5ex}l^e_{n_c, k} (\bar{\mathbf{e}}_{n_c}, \bar{\mathbf{x}}_k) \! \propto \! \prod_{u \in \mathcal{U}} \! \! \left \{ q^x_{c,u,k}(x_{u,k}) \right \}^{\frac{1}{N_c}} \! \! \prod_{u \in \mathcal{U}}  q^e_{n_c,u,k}(e_{n_c,u}). \!\!\!
    \end{align}
    \end{subequations}

    To solve the KL minimization problem, the alternating optimization algorithm 
    \cite{2017jain_optimization_tutorial} 
    is utilized, where a target parameter is optimized while the other parameters are fixed.
    In what follows, the estimation methods for $ \{\bm{\pi}^{q,x}_{c,u,k}, \bm{\pi}^{b,x}_{u,k}\} $ and $ \{\bm{\pi}^{b,e}_{n,u}$, $\bm{\pi}^{q,e}_{n,u,k} \}$ are described in Section \ref{subsec:EP_x} and \ref{subsec:EP_e}, respectively.

    \vspace{-0.3em}
    \subsection{EP for Data Estimation}
    \label{subsec:EP_x}
     
    \subsubsection{Update $\bm{\pi}^{q,x}_{c,u,k}$}
    While the parameter $\bm{\pi}^{q,x}_{c,u,k}$ in $q^x_{c,u,k}(x_{u,k})$ is updated, the other parameters
    $\mathbf{\Pi} \setminus \{\bm{\pi}^{q,x}_{c,u,k} \}$ are fixed as the tentative estimated values, that is, the KL minimization problem for $\bm{\pi}^{q,x}_{c,u,k}$ is formulated as
    \begin{align}
        \label{eq:min_qx}
        \underset{\boldsymbol{\pi}^{q,x}_{c,u,k}}{\mathrm{minimize}}\ 
        \mathrm{KL} \left(
        \hat{p}_{c,k}^{q,x}(\mathbf{E}, \mathbf{X} | \mathbf{Y}) \|
        g(\mathbf{E}, \mathbf{X} | \mathbf{Y}) 
        \right),
    \end{align}
    where $\hat{p}_{c,k}^{q,x}(\mathbf{E}, \mathbf{X} | \mathbf{Y}) $ is the target distribution for $\bm{\pi}^{q,x}_{c,u,k}$, which is designed using $l^x_{c, k} (\mathbf{E}_{c}, \bar{\mathbf{x}}_k)$ in \eqref{eq:likelihood_apx_x} as
    \begin{align}
        \hat{p}_{c,k}^{q,x}(& \mathbf{E}, \mathbf{X} | \mathbf{Y}) 
        = C_{c,k}^{q,x}\ 
        p(\mathbf{y}_{c,k} | \mathbf{E}_c, \bar{\mathbf{x}}_k)  \nonumber \\
        & \prod_{(c^\prime,k^\prime) \in \mathcal{C} \times \mathcal{K} \setminus (c,k)}
        \underbrace{l^x_{c^\prime, k^\prime} (\mathbf{E}_{c^\prime}, \mathbf{x}_{k^\prime}) }_{\simeq p(\mathbf{y}_{c^\prime,k} | \mathbf{E}_{c^\prime}, \mathbf{x}_{k^\prime})  }
        \underbrace{B^x(\mathbf{X}) B^e(\mathbf{E})}_{\simeq p(\mathbf{X}) p(\mathbf{E};\mathbf{\Theta})},
    \end{align}
    where $C_{c,k}^{q,x}$ is a normalizing constant.
    
    Let $\mathcal{L}^{q,x}_{c,u,k} (\bm{\pi}^{q,x}_{c,u,k}) \triangleq \mathrm{KL} \left( \hat{p}_{c,k}^{q,x}(\mathbf{E}, \mathbf{X} | \mathbf{Y}) \| g(\mathbf{E},\mathbf{X} | \mathbf{Y}) \right)$ denote the objective function in \eqref{eq:min_qx}, resorting to
    \begin{align}
        \mathcal{L}^{q,x}_{c,u,k} 
        = \! \ln Z_g \! - \! \mathbb{E}_{\hat{p}^{q,x}_{c,k} (x_{u,k} | \mathbf{Y})} \! \left [ \ln q^x_{c,u,k}(x_{u,k}) \right ] + \mathrm{const}.
    \end{align}
    
    Since the objective function $\mathcal{L}^{q,x}_{c,u,k}(\boldsymbol{\pi}_{c,u,k}^{q,x})$ is convex with respect to $\bm{\pi}_{c,u,k}^{q,x}$, the necessary and sufficient condition for the global optimal
    , \textit{i.e.,} $\partial \mathcal{L}^{q,x}_{c.u,k} / \partial \boldsymbol{\pi}_{c,u,k}^{q,x}= \mathbf{0}$, is equivalent to 
    \begin{align}
        \label{eq:proj_qx}
        g(x_{u,k} | \mathbf{Y}) = \mathrm{proj}_{\mathbf{\Phi}} \left[ \hat{p}^{q,x}_{c,k} (x_{u,k}| \mathbf{Y}) \right],
    \end{align}
    where 
    $\mathrm{proj}_{\mathbf{\Phi}} \left[ p(x) \right] \triangleq \mathcal{CN} \left (\mathbb{E}_{p(x)}[x],\ \mathbb{V}_{p(x)}[x] \right )$ 
    is the projection operator onto Gaussian distribution set $\mathbf{\Phi}$, which indicates the moment matching, \textit{i.e.,} the first and second moments of distribution $p(x)$ matches those of the target distribution.
    
    The marginalized approximate posterior $g(x_{u,k} | \mathbf{Y}) = \int_{\mathbf{E}, \mathbf{X} \setminus x_{u,k}} g(\mathbf{E}, \mathbf{X} | \mathbf{Y})$ in \eqref{eq:proj_qx} is written as
    \begin{align}
        g(x_{u,k} | \mathbf{Y}) 
        \label{eq:g_qx_2}
        & \propto  q^x_{c, u,k}(x_{u,k}) v^x_{c, u,k}(x_{u,k}),
    \end{align}
    with $v^x_{c, u,k}(x_{u,k})\triangleq b_{u,k}^x(x_{u,k}) \prod_{c^\prime \in \mathcal{C} \setminus c} q^x_{c^\prime, u,k}(x_{u,k})$, which can also be represented as
    \begin{align}
        v^x_{c, u,k}(x_{u,k}) 
        \label{eq:vx}
        &= C^{v,x}_{c,u,k} \exp \left ( - |x_{u,k} - \hat{x}^{v}_{c,u,k} |^2 / \xi^{v,x}_{c,u,k} \right ),
    \end{align}
    with the normalizing constant $C^{v,x}_{c,u,k}$.
    
    The marginalized target distribution $\hat{p}^{q,x}_{c,k} (x_{u,k}| \mathbf{Y}) = \int_{\mathbf{E}, \mathbf{X} \setminus x_{u,k}} \hat{p}^{q,x}_{c,k}(\mathbf{E}, \mathbf{X} | \mathbf{Y})$ in \eqref{eq:proj_qx} is written as 
    \begin{align}
        \label{eq:p_tg_qx}
        \hat{p}^{q,x}_{c,k} (x_{u,k}| \mathbf{Y})
        &= \bar{p}^x_{c,u,k} (\mathbf{y}_{c,k} | x_{u,k}) v^x_{c,u,k}(x_{u,k}),
    \end{align}
    where $\bar{p}^x_{c,u,k} (\mathbf{y}_{c,k} | x_{u,k})$ is the conditional probability distribution defined in \eqref{eq:px_cond} at the top of next page along with $v^e_{n_c,u, k} (e_{n_c,u}) $, and $\tilde{v}^e_{c,u,k} (\mathbf{e}_{c,u})$, with $\bar{C}^x_{c,u,k}$ being the normalizing constant and
    \setcounter{equation}{37}
    \begin{align}
        \hat{\mathbf{e}}^v_{c,u,k} &=  \begin{bmatrix} \hat{e}^v_{n_{c(1)},u,k}, \hat{e}^v_{n_{c(2)},u,k}, \ldots, \hat{e}^v_{n_{c(N_c)},u,k} \end{bmatrix}^\mathrm{T}  \in \mathbb{C}^{N_c \times 1},  \nonumber \\
        \mathbf{\Xi}^{v,e}_{c,u,k} \!\! &= \! \mathrm{diag} \left( \!\xi^{v,e}_{n_{c(1)},u,k}, \xi^{v,e}_{n_{c(2)},u,k}, \ldots, \xi^{v,e}_{n_{c(N_c)},u,k}
        \! \right)\in \mathbb{R}^{N_c \times N_c}. \nonumber
    \end{align}
    
    From the conditional distribution $\bar{p}^x_{c,u,k} (\mathbf{y}_{c,k} | x_{u,k})$ in \eqref{eq:px_cond}, 
    the mean
    $\tilde{\mathbf{y}}^x_{c,u,k} \triangleq \mathbb{E}_{ \bar{p}^x_{c,u,k} (\mathbf{y}_{c,k} | x_{u,k}) }  \left [  \mathbf{y}_{c,k} \right]$ 
    and covariance
    $\mathbf{\Omega}^{x}_{c,k} \triangleq \mathbb{E}_{ \bar{p}^x_{c,u,k} (\mathbf{y}_{c,k} | x_{u,k}) } \left [  (\mathbf{y}_{c,k} - \tilde{\mathbf{y}}^x_{c,u,k} ) (\mathbf{y}_{c,k} - \tilde{\mathbf{y}}^x_{c,u,k} )^\mathrm{H} \right]$,
    can be calculated as
    \begin{subequations}
    \begin{align}
        \label{eq:mean_IC}
        \tilde{\mathbf{y}}^x_{c,u,k} 
        &= \mathbf{y}_{c,k} - \sum_{u^\prime \in \mathcal{U} \setminus u} \ \hat{\mathbf{h}}^v_{c,u^\prime,k} \hat{x}^v_{c, u^\prime, k},  \\
        \label{eq:cov_IC}
        \mathbf{\Omega}^{x}_{c,k} 
        &= \sum_{u^\prime \in \mathcal{U}}  
        \Big \{ \xi^{v,x}_{c,u^\prime,k} 
        \hat{\mathbf{h}}^{v}_{c,u^\prime,k} \hat{\mathbf{h}}^{v \mathrm{H}}_{c,u^\prime,k}   \nonumber \\
        & \qquad  + (\xi^{v,x}_{c,u^\prime,k} + |\hat{x}^{v}_{c,u^\prime,k}|^2) \mathbf{\Xi}^{v,e}_{c,u^\prime,k} 
        \Big \} +\sigma^2 \mathbf{I}_{N_c}, 
    \end{align}
    \end{subequations}
    with $\hat{\mathbf{h}}^v_{c,u^\prime,k} \triangleq \hat{\mathbf{e}}^v_{c,u^\prime,k} +\hat{\mathbf{s}}_{c,u^\prime}$.

    Substituting \eqref{eq:g_qx_2} and \eqref{eq:p_tg_qx} into \eqref{eq:proj_qx}, the approximate function $q_{c,u,k}^x(x_{u,k})$ can be obtained as 
    \begin{align}
        \label{eq:qx_proj_v}
        q^x_{c,u,k}(x_{u,k}) 
        \propto  \frac{\mathrm{proj}_\mathbf{\Phi} \left [ \bar{p}^x_{c,u,k} (\mathbf{y}_{c,k} | x_{u,k}) v^x_{c,u,k}(x_{u,k}) \right ]}{v^x_{c, u,k}(x_{u,k})}.
    \end{align}
    
    Under large system conditions with \ac{CLT}, the conditional distribution can be approximated as $\bar{p}^x_{c,u,k} (\mathbf{y}_{c,k} | x_{u,k}) \simeq \mathcal{CN}(\tilde{\mathbf{y}}^x_{c,u,k}, \mathbf{\Omega}^{x}_{c,k} )$.
    Thus, the approximate function can be expressed as $q^x_{c,u,k}(x_{u,k}) \propto \bar{p}^x_{c,u,k} (\mathbf{y}_{c,k} | x_{u,k})$
    with the mean and variance calculated as
    \begin{subequations}
    \begin{align}
        \label{eq:xq}
        \hat{x}^{q}_{c,u,k} &= 
        \hat{\mathbf{h}}^{v \mathrm{H}}_{c,u,k}
        (\mathbf{\Omega}_{c,k}^{x})^{-1} \tilde{\mathbf{y}}^x_{c,u,k} / \gamma^x_{c,u,k}, \\
        \label{eq:xiq}
        \xi^{q,x}_{c,u,k} &= 1 / \gamma^x_{c,u,k} - \xi^{v,x}_{c,u,k},
    \end{align}
    \end{subequations}
    with $\gamma^{x}_{c,u,k} =  \hat{\mathbf{h}}^{v \mathrm{H}}_{c,u,k} (\mathbf{\Omega}^{x}_{c,k})^{-1} \hat{\mathbf{h}}^{v}_{c,u,k}$.
    The calculation of $\tilde{\mathbf{y}}^x_{c,u,k} $ in \eqref{eq:mean_IC} corresponds to a \ac{Soft-IC} \cite{2023Ito_AoA} 
    using data replicas 
    $\{ \hat{x}^v_{c,u^\prime,k} \}_{u^\prime \in \mathcal{U} \setminus u}$ and channel replicas 
    $\{ \hat{\mathbf{h}}^v_{c,u^\prime,k} \}_{u^\prime \in \mathcal{U}}$.

    Unlike the conventional MRC-based detections \cite{2019Yan_bilinear,2019Chen_bilinear_cluster,2023Ito_AoA}, the LMMSE-based detection for each sub-array in \eqref{eq:xq} can deal with the correlation between the leaked energy in the beam-domain owing to the whitening operation by $(\mathbf{\Omega}_{c,k}^{x})^{-1}$.
    The correlation within the $c$-th sub-array, corresponding to the off-diagonal elements of $\mathbf{\Omega}^{x}_{c,k}$, increases as the leaked energy increases because the off-diagonal elements is computed by the weighted summation of the outer product of channel replicas $\hat{\mathbf{h}}^{v}_{c,u^\prime,k} \hat{\mathbf{h}}^{v \mathrm{H}}_{c,u^\prime,k}$ in~\eqref{eq:cov_IC}.
    %
    As illustrated in Fig.~\ref{fig:NF_DFT}, the energy leakage effects become prominent as the distance between the BS and the UE decrease due to the near-field effects.
    Therefore, when the number of antennas per sub-array $N_c$ is small, including conventional MRC-based detections ($N_c=1$), each sub-array cannot address the correlation caused by the energy leakage.
    The number of antennas per sub-array $N_c$ should be selected by considering the trade-off between data detection performance and its computational complexity, which are evaluated through numerical simulations in Section~\ref{sec:simulation}.
    
    \subsubsection{Update $\bm{\pi}^{b,x}_{u,k}$}

    The KL minimization problem for $\bm{\pi}^{b,x}_{u,k}$ is formulated as
    \begin{align}
        \label{eq:min_bx}
        \underset{\boldsymbol{\pi}^{b,x}_{u,k}}{\mathrm{minimize}}\ 
        \mathrm{KL} \left(
        \hat{p}_{u,k}^{b,x}(\mathbf{E}, \mathbf{X} | \mathbf{Y}) \|
        g(\mathbf{E}, \mathbf{X} | \mathbf{Y})
        \right),
    \end{align}
    where $\hat{p}_{u,k}^{b,x}(\mathbf{E}, \mathbf{X} | \mathbf{Y}) $ is the target distribution defined as
    \begin{align}
        &\hat{p}_{u,k}^{b,x}(\mathbf{E}, \mathbf{X} | \mathbf{Y})  = \ C_{u,k}^{b,x} \ p(x_{u,k})  \nonumber \\
        & 
        \prod_{(u^\prime,k^\prime) \in \mathcal{U} \times \mathcal{K} \setminus (u,k)} \underbrace{b_{u^\prime,k^\prime}^x (x_{u^\prime,k^\prime})}_{\simeq p(x_{u^\prime, k^\prime})}
        \underbrace{Q^x(\mathbf{X})  Q^e(\mathbf{E})}_{ \simeq p(\mathbf{Y} | \mathbf{E}, \mathbf{X})}
        \underbrace{B^e(\mathbf{E})}_{\simeq p(\mathbf{E};\mathbf{\Theta})} 
    \end{align}
    where $C_{u,k}^{b,x}$ is a normalizing constant.
    
    Similar to the derivation of \eqref{eq:proj_qx}, the optimal condition for $\bm{\pi}^{b,x}_{u,k}$ is derived as 
    \begin{align}
        \label{eq:gx_bx}
        g(x_{u,k} | \mathbf{Y}) = \mathrm{proj}_{\mathbf{\Phi}} \left[ \hat{p}^{b,x}_{u,k} (x_{u,k}| \mathbf{Y}) \right],
    \end{align}
    where 
    $\hat{p}^{b,x}_{u,k} (x_{u,k}| \mathbf{Y}) = \int_{\mathbf{E}, \mathbf{X} \setminus x_{u,k}} \hat{p}^{b,x}_{u,k}(\mathbf{E}, \mathbf{X} | \mathbf{Y})$ is the marginalized target distribution calculated as
    \begin{align}
        \label{eq:p_hat_bx}
        \hat{p}^{b,x}_{u,k} (x_{u,k}| \mathbf{Y})
        & \propto\ p(x_{u,k}) \prod_{c^\prime \in \mathcal{C}} q^x_{c^\prime,u,k}(x_{u,k}),
    \end{align}
    with the approximate function multiplied over the sub-array direction,
    $\prod_{c^\prime \in \mathcal{C}} q^x_{c^\prime,u,k}(x_{u,k})$, calculated as 
    \begin{align}
        \label{eq:qx_prod}
        \prod_{c^\prime \in \mathcal{C}} q^x_{c^\prime,u,k}(x_{u,k})
        \propto 
        \exp \left ( -| x_{u,k} - \hat{x}^q_{u,k} |^2 / \xi^{q,x}_{u,k} \right), 
    \end{align}
    \begin{align}
        \label{eq:qx_prod_mean_vari}
        \hat{x}_{u,k}^{q} = \xi_{u,k}^{q,x} \left( \sum_{c^\prime \in \mathcal{C}}\frac{\hat{x}_{c^\prime,u,k}^{q}}{\xi_{c^\prime,u,k}^{q,x}}  \right),\ 
        \xi_{u,k}^{q,x} = \left( \sum_{c^\prime \in \mathcal{C}}\frac{1}{\xi_{c^\prime,u,k}^{q,x}}  \right)^{-1} \! \! \! \! \!.
    \end{align}
    
    Note that combining the mean $\{\hat{x}^q_{c,u,k}\}_{c \in \mathcal{C}}$ and variance $\{ \xi^{q,x}_{c,u,k} \}_{c \in \mathcal{C}}$ over the sub-array direction $c \in \mathcal{C}$, as written in \eqref{eq:qx_prod_mean_vari}, leads to further improvements for data detection owing to the spatial diversity.
    Substituting \eqref{eq:p_hat_bx} into \eqref{eq:gx_bx}, the approximate posterior $g(x_{u,k}|\mathbf{Y})$ can be written as
    \begin{align}
        g(x_{u,k} | \mathbf{Y})
        \label{eq:g_qx_3}
        & \propto \mathrm{proj_\mathbf{\Phi}}  \left [ p(x_{u,k}) \prod_{c^\prime \in \mathcal{C}} q^x_{c^\prime,u,k}(x_{u,k}) \right ].
    \end{align}
    
    The approximate posterior mean $\hat{x}_{u,k}$ and variance $\xi^{x}_{u,k}$ of $g(x_{u,k} | \mathbf{Y})$ can be derived using the MMSE denoiser function $\eta(\cdot)$ \cite{2022zou_AMP_tutorial}, which is designed based on the prior for \ac{QAM} constellation $p(x_{u,k})$ in \eqref{eq:prior_x}.
    Then, the posterior mean and variance is expressed as
    $\hat{x}_{u,k} = \eta(\hat{x}^q_{u,k}, \xi^{q,x}_{u,k}) \triangleq \mathbb{E}_{g(x_{u,k}|\mathbf{Y})} [x_{u,k}]$ and 
    $\xi^{x}_{u,k} = \xi^{q,x}_{u,k} \frac{\partial \eta(\hat{x}^q_{u,k}, \xi^{q,x}_{u,k})}{\partial \hat{x}^q_{u,k}} $, which can be calculated as 
    \begin{subequations}
    \begin{align}
        \label{eq:x_denoise}
        & \! \! \hat{x}_{u,k} = C^{g,x}_{u,k} \sum_{\mathcal{X}_q \in \mathcal{X}}  \mathcal{X}_q \exp
        \left ( -| \mathcal{X}_q - \hat{x}^q_{u,k} |^2 / \xi^{q,x}_{u,k} \right),  \\ 
        \label{eq:xi_denoise}
        & \! \! \xi^{x}_{u,k} \! = \! 
        C^{g,x}_{u,k}
        \! \! \sum_{\mathcal{X}_q \in \mathcal{X}} \! \! |\mathcal{X}_q|^2 \exp \!
        \left ( \! -| \mathcal{X}_q \! - \! \hat{x}^q_{u,k} |^2 / \xi^{q,x}_{u,k} \! \right) \! - \! |\hat{x}_{u,k}|^2 \! \! , \!
    \end{align}
    \end{subequations}
    with 
    $(C^{g,x}_{u,k})^{-1} = \sum_{\mathcal{X}_q \in \mathcal{X}}  \exp \left ( -| \mathcal{X}_q - \hat{x}^q_{u,k} |^2 / \xi^{q,x}_{u,k} \right).$

    From \eqref{eq:g_qx_2}, $v_{c,u,k}^x (x_{u,k})$ can be updated as
    \begin{align}
        \label{eq:vx_gq}
        v^x_{c,u, k} (x_{u,k}) \propto\  g(x_{u,k} | \mathbf{Y}) / q^x_{c,u,k}(x_{u,k}),
    \end{align}
    where the associated mean and variance are given by
    \begin{align}
        \label{eq:vx_mean_vari}
        \hat{x}^{v}_{c,u,k} \!=\! \xi^{v,x}_{c,u,k} \left(  \frac{\hat{x}_{u,k}}{\xi^{x}_{u,k}} \!-\! \frac{\hat{x}^{q}_{c,u,k}}{\xi^{q,x}_{c,u,k}}\right)\!,\ 
        \xi^{v,x}_{c,u,k} \!=\! \left( \! \frac{1}{\xi^{x}_{u,k}} \!-\! \frac{1}{\xi^{q,x}_{c,u,k}} \! \right)^{-1} \!.
    \end{align}

    As shown in the \ac{Soft-IC} process in \eqref{eq:mean_IC}-\eqref{eq:cov_IC}, $\hat{x}^{v}_{c,u,k}$ and $\xi^{v,x}_{c,u,k}$ in \eqref{eq:vx_mean_vari} are used as soft replicas instead of $\hat{x}_{u,k}$ and $\xi^{x}_{u,k}$ in \eqref{eq:x_denoise}-\eqref{eq:xi_denoise} in order to suppress the self-noise feedback in the algorithm iterations \cite{2022Tamaki_GAMP_feedback}.
    In conventional JCDE algorithms \cite{2024Ito_BiGaBP,2023Ito_AoA}, the self-feedback suppression is performed before the denoising process in \eqref{eq:x_denoise}-\eqref{eq:xi_denoise} by generating antenna-wise extrinsic values based on \ac{BP} rules.
    Hence, the complexity of the denoising process is $\mathcal{O}(QNUK_\mathrm{d})$.
    In contrast, the proposed method can reduce the complexity in the  denoising process as $\mathcal{O}(QUK_\mathrm{d})$, since the extrinsic values $\hat{x}^{v}_{c,u,k}$ and $\xi^{v,x}_{c,u,k}$ are generated after the denoising process in \eqref{eq:vx_mean_vari}.
     
    Finally, from \eqref{eq:g_qx_2}, $b^{x}_{u,k}(x_{u,k})$ can be updated as
    \begin{align}
        \label{eq:bx_update}
        b^x_{u,k}(x_{u,k}) \ &\propto\ g(x_{u,k} | \mathbf{Y}_{u,k}) \Big / \prod_{c^\prime \in \mathcal{C}} q^x_{c^\prime,u,k}(x_{u,k}),
    \end{align}
    with the mean $\hat{x}^{b}_{u,k}$ and variance $\xi^{b,x}_{u,k}$ of $b^x_{u,k}(x_{u,k})$ being
    \begin{align}
        \label{eq:xb_mean_vari}
        &\hat{x}^{b}_{u,k} \!=\! \xi^{b,x}_{u,k} \left(  \frac{\hat{x}_{u,k}}{\xi^{x}_{u,k}} - \frac{\hat{x}^{q}_{u,k}}{\xi^{q,x}_{u,k}} \right), \! \! \!
        &\xi^{b,x}_{u,k} \!=\! \left(  \frac{1}{\xi^{x}_{u,k}} - \frac{1}{\xi^{q,x}_{u,k}} \right)^{-1} \! \! \! \! \! \!.
    \end{align}

    \subsection{EP for Residual Channel Error Estimation} 
    \label{subsec:EP_e}

    \subsubsection{Update $\bm{\pi}^{q,e}_{n_c,u,k}$}
    For $\bm{\pi}^{q,e}_{n_c,u,k}$, we minimize
    \begin{align}
        \label{eq:min_qe}
        \underset{\boldsymbol{\pi}^{q,e}_{n_c,u,k}}{\mathrm{minimize}}\ 
        \mathrm{KL} \left( \hat{p}_{n_c,k}^{q,e}(\mathbf{E}, \mathbf{X} | \mathbf{Y}) \| g(\mathbf{E}, \mathbf{X} | \mathbf{Y}) \right),
    \end{align}
    where $\hat{p}_{n_c,k}^{q,e}(\mathbf{E}, \mathbf{X} | \mathbf{Y}) $ is the target distribution designed as
    \begin{align}
        &\hat{p}_{n_c,k}^{q,e}(\mathbf{E}, \mathbf{X} | \mathbf{Y}) 
        =
        C_{n_c,k}^{q,e} \ 
        p(y_{n_c,k} | \bar{\mathbf{e}}_{n_c}, \bar{\mathbf{x}}_k) 
     \nonumber \\
        & 
        \prod_{(n_c^\prime, k^\prime) \in \mathcal{N} \times \mathcal{K}\setminus (n_c,k)}
        \underbrace{
            l^e_{n_c^\prime, k^\prime}(\bar{\mathbf{e}}_{n_c^\prime}, \mathbf{x}_{k^\prime} )
        }_{\simeq p(y_{n_c^\prime, k^\prime} | \bar{\mathbf{e}}_{n_c^\prime}, \mathbf{x}_{k^\prime})}
        \underbrace{B^x(\mathbf{X})}_{\simeq p(\mathbf{X})}
        \underbrace{B^e(\mathbf{E})}_{\simeq p(\mathbf{E};\mathbf{\Theta})},
    \end{align}
    where $C_{n_c,k}^{q,e} $ is a normalizing constant.
    
    Through the same procedure as the derivation of $q^x_{c,u,k}(x_{u,k})$ in \eqref{eq:qx_proj_v}, the mean and variance of approximate function 
    $q^e_{n_c,u,k}(e_{n_c,u})$
    are obtained as 
    \begin{align}
        \label{eq:qe_mean_vari}
        \hat{e}^q_{n_c,u,k} = \frac{\hat{x}^{w \ast}_{n_c,u,k} \tilde{y}^e_{n_c,u,k}}{|\hat{x}_{n_c,u,k}^w|^2},\ 
        \xi^{q,e}_{n_c,u,k} = \frac{\phi^e_{n_c,u,k}}{|\hat{x}_{n_c,u,k}^w|^2},
    \end{align}
    with 
    \begin{subequations}
    \begin{align}
        \label{eq:mean_IC_e}
        \tilde{y}^e_{n_c,u,k} 
        &= y_{n_c,k} \! -  \! \sum_{u^\prime \in \mathcal{U} \setminus u} \hat{x}^w_{c,u^\prime,k} \hat{e}^v_{n_c, u^\prime,k} - \sum_{u^\prime \in \mathcal{U}} \hat{x}^w_{n_c,u^\prime,k} \hat{s}_{n_c,u^\prime}, \\
        \label{eq:cov_IC_e}
        \phi^e_{n_c,u,k} 
        & = \sum_{u^\prime \in \mathcal{U}} \left( 
        |\hat{e}^v_{n_c,u^\prime,k}|^2 + |\hat{s}_{n_c,u^\prime}|^2 + \xi^{v,e}_{c,u^\prime,k}
        \right) \xi^{w,x}_{n_c,u^\prime,k} \nonumber \\
        & \quad + \sum_{u^\prime \in \mathcal{U} \setminus u} \xi^{v,e}_{n_c,u^\prime,k} |\hat{x}^w_{c,u^\prime,k}|^2  + \sigma^2, \\
        \label{eq:wx_mean}
        \hat{x}^{w}_{c,u,k} &=
        \xi^{w,x}_{c,u,k} \left( \hat{x}^x_{u,k} (\xi^{x}_{u,k})^{-1} - \hat{x}^{q}_{c,u,k} (N_c \xi^{q,x}_{c,u,k})^{-1} \! \right), \! \! \\
        \label{eq:wx_vari}
        \xi^{w,x}_{c,u,k} &= \left( (\xi^x_{u,k})^{-1} - (N_c \xi^{q,x}_{c,u,k})^{-1} \right)^{-1}.
    \end{align}
    \end{subequations}

    \subsubsection{Update $\bm{\pi}^{b,e}_{n_c,u}$} 
    For $\bm{\pi}^{b,e}_{n_c,u}$, we have
    \begin{align}
        \label{eq:min_be}
        \underset{\boldsymbol{\pi}^{b,e}_{n_c,u}}{\mathrm{minimize}}\ 
        \mathrm{KL} \left( \hat{p}_{n_c,u}^{b,e}(\mathbf{E}, \mathbf{X} | \mathbf{Y}) \|
        g(\mathbf{E}, \mathbf{X} | \mathbf{Y}) \right),
    \end{align}
    where $\hat{p}_{n_c,u}^{b,e}(\mathbf{E}, \mathbf{X} | \mathbf{Y})$ is the target distribution designed as
    \begin{align}
        & \hat{p}_{n_c,u}^{b,e}(\mathbf{E}, \mathbf{X} | \mathbf{Y}) = 
        C_{n_c,u}^{b,e} \ p(e_{n_c,u};\mathbf{\Theta}) \nonumber \\
        & \! \! \! \! \cdot \! \! \! \! \prod_{(n_c^\prime,u^\prime) \in \mathcal{N} \times \mathcal{U} \setminus (n_c,u)} 
        \underbrace{ b_{n_c^\prime,u^\prime}^e (e_{n_c^\prime,u^\prime})}_{\simeq p(e_{n_c, u} ; \mathbf{\Theta} )}
        \underbrace{ Q^x(\mathbf{X}) Q^e(\mathbf{E}) }_{ \simeq p( \mathbf{Y} | \mathbf{E}, \mathbf{X})}
        \underbrace{B^x(\mathbf{X})}_{\simeq p(\mathbf{X})},
    \end{align}
    where $C_{n_c,u}^{b,e}$ is a normalizing constant.

    Following the same methodology used to derive $g(x_{u,k}|\mathbf{Y})$ in \eqref{eq:g_qx_3}, the approximate posterior $g(e_{n_c,u}| \mathbf{Y})$
    are derived as
    \begin{align}
        \label{eq:p_tar_be}
        g(e_{n_c,u}| \mathbf{Y}) \! &\propto \mathrm{proj}_\mathbf{\Phi} \! \left [
        p(e_{n_c,u};\mathbf{\Theta})   \prod_{k^\prime \in \mathcal{K}} q_{n_c,u,k^\prime}^e(e_{n_c,u})
        \right] \! ,
    \end{align}
    where the mean and variance of $g(e_{n_c,u}| \mathbf{Y})$ can be calculated based on the prior distribution $p(e_{n_c,u}; \mathbf{\Theta})$ in \eqref{eq:prior_e} as
    \begin{align}
        \label{eq:e_denoise}
        \hat{e}_{n_c,u} = \frac{\sigma^e_{n_c,u} \hat{e}^{q}_{n_c,u}}{ \sigma^e_{n_c,u} + \xi^{q,e}_{n_c,u} }, \ 
        \xi^{e}_{n_c,u} = \left( \frac{1}{\sigma^{e}_{n_c,u}} + \frac{1}{\xi^{q,e}_{n_c,u}} \right)^{-1},
    \end{align}
    with
    \begin{align}
        \label{eq:qe_prod_mean_vari}
        \hat{e}_{n_c,u}^{q} \! = \! \xi_{n_c, u}^{q,e} \! \left( \sum_{k^\prime \in \mathcal{K}}\frac{\hat{e}_{n_c,u,k^\prime}^{q}}{\xi_{n_v,u,k^\prime}^{q,e}} \! \right),
        \xi_{n_c, u}^{q,e} \! \! = \! \left( \sum_{k^\prime \in \mathcal{K}}\frac{1}{\xi_{n_c,u,k^\prime}^{q,e}} \! \right)^{-1} \! \! \! \! \! \! \! \!.
    \end{align}

    Similarly, the approximate function $v^e_{n_c,u, k} (e_{n_c,u})$ can be derived in the same manner as \eqref{eq:vx_gq}:
    \begin{align}
        \label{eq:v_e_gq}
        v^e_{n_c,u, k} (e_{n_c,u}) \propto g(e_{n_c,u}|\mathbf{Y}) / q_{n_c,u,k}^e(e_{n_c,u}),
    \end{align}
    from which the mean and variance are respectively given by
    \begin{subequations}
    \begin{align}
        \label{eq:ve_mean}
        & \! \hat{e}^{v}_{n_c,u,k} \!=\! \xi^{v,e}_{n_c,u,k}\left(\! (\xi^{e}_{n_c,u})^{-1} \hat{e}_{n_c,u} \!-\!  (\xi^{q,e}_{n_c,u,k})^{-1} \hat{e}^q_{n_c,u,k} \! \right), \! \\ 
        \label{eq:ve_vari}
        & \! \xi^{v,e}_{n_c,u,k} = \left( (\xi^{e}_{n_c,u})^{-1} - (\xi^{q,e}_{n_c,u,k})^{-1} \right)^{-1} .
    \end{align}
    \end{subequations}
    
    Finally, the approximate function $b^e_{n_c,u}(e_{n_c,u})$ is obtained in a similar way as the derivation of \eqref{eq:bx_update} as
    \begin{align}
        \label{eq:be_up}
        b^e_{n_c,u}(e_{n_c,u}) 
        &\propto  g(e_{n_c,u} | \mathbf{Y}) \Big / \prod_{k^\prime \in \mathcal{K}} q_{n_c,u,k^\prime}^e(e_{n_c,u}),
    \end{align}
    with the mean and variance of $b^e_{n_c,u}(e_{n_c,u})$ being
    \begin{align}
        \label{eq:be_mean_vari}
        \hat{e}^{b}_{n_c,u} \!=\! \xi^{b,e}_{n_c,u} \left(  \frac{\hat{e}_{n_c,u}}{\xi^{e}_{n_c,u}} - \frac{\hat{e}^{q}_{n_c,u}}{\xi^{q,e}_{n_c,u}} \right),\ 
        \xi^{b,e}_{n_c,u} \!=\! \left(  \frac{1}{\xi^{e}_{n_c,u}} - \frac{1}{\xi^{q,e}_{n_c,u}} \right)^{-1} \!\!.
    \end{align}

    \begin{figure*}[t]
        \centering
        \includegraphics[width=\linewidth]{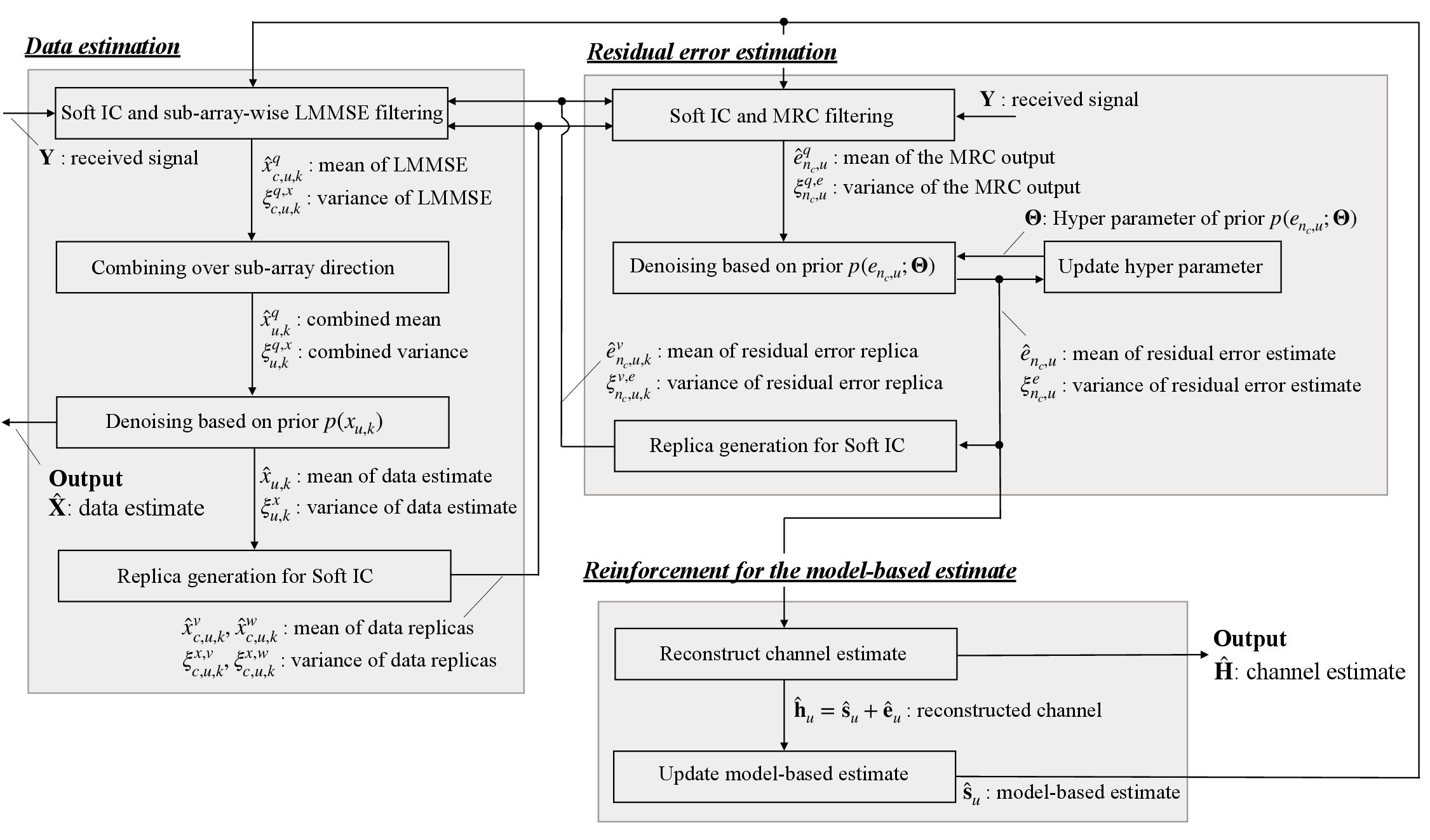}
        \caption{Framework of the proposed JCDE algorithm.}
        \label{fig:alg_flow}
    \end{figure*}

    \subsection{Expectation Maximization for Hyperparameter Learning}
    \label{subsec:EM}   
    In this section, we describe the estimation method for hyperparameter set $\mathbf{\Theta}$ via the \ac{EM} algorithm corresponding to M-step in \eqref{eq:EM_M}.
    Using the approximate posterior $g^{(t)}(\mathbf{E}, \mathbf{X} | \mathbf{Y})$ at the $t$-th step as descrbed in Section \ref{subsec:EP_x} and \ref{subsec:EP_e}, the \ac{ELBO} $\mathcal{F} \left( \mathbf{\Theta}, \mathbf{\Theta}^{(t)} \right)$ in \eqref{eq:EM_M} can be approximated as 
    \begin{align}
        \mathcal{F}(\mathbf{\Theta}, & \mathbf{\Theta}^{(t)})
        \simeq
        -\sum_{n \in \mathcal{N}} \sum_{u \in \mathcal{U}} 
        \Big \{ \ln \sigma^e_{n_c,u}  + \nonumber \\
        & (\sigma^e_{n_c,u})^{-1} \mathbb{E}_{g^{(t)}(e_{n_c,u}| \mathbf{Y}) }  
        \left [|e_{n_c,u}|^2 \right ] \Big \} 
        + \mathrm{const}.
    \end{align}
    
    Since the \ac{ELBO} $\mathcal{F}(\mathbf{\Theta}, \mathbf{\Theta}^{(t)})$ is concave for $(\sigma^e_{n,u})^{-1}$, 
    the maximization problem in \eqref{eq:EM_M} can be solved by the first-order necessary and sufficient condition 
    $\partial \mathcal{F}(\mathbf{\Theta}, \mathbf{\Theta}^{(t)}) / \partial \left(\sigma_{n,u}^e \right)^{-1} = \mathbf{0}$
    , which derives the optimal variance $\sigma^{e(t+1)}_{n,u}$ at the $t$-th step as 
    \begin{align}
        \label{eq:sigma_EM}
        \sigma^{e(t+1)}_{n_c,u} &= \mathbb{E}_{g^{(t)}(e_{n_c,u}| \mathbf{Y}) }   \left [|e_{n_c,u}|^2 \right ] 
        = |\hat{e}^{(t)}_{n_c,u}|^2 + \xi^{e(t)}_{n_c,u},
    \end{align}
    where $\hat{e}^{(t)}_{n_c,u}$ and $\xi^{e(t)}_{n_c,u}$ are the approximate posterior mean and variance at the $t$-th step, calculated in \eqref{eq:e_denoise}.

    \subsection{Reinforcement for the Model-Based Estimate}
    \label{subsec:MB}

    To further improve the convergence performance for the EP algorithm, 
    we update the model-based estimate $\hat{\mathbf{S}}$ in each iteration.
    Using the estimated residual channel error $\hat{\mathbf{E}}^{(t-1)} \triangleq \mathbb{E}_{g^{(t-1)}(\mathbf{E}, \mathbf{X}| \mathbf{Y})} [\mathbf{E}]$ at the $(t-1)$-th iteration, the channel estimate for the $u$-th UE can be reconstructed as 
    \begin{align}
        \label{eq:H_est_S}
        \hat{\mathbf{h}}^{(t-1)}_u = \hat{\mathbf{s}}^{(t-1)}_u + \hat{\mathbf{e}}^{(t-1)}_u.
    \end{align}
    
    The model-based estimate at the $t$-th iteration $\hat{\mathbf{s}}_u^{(t)}$
    is updated with the channel estimate at the previous iteration $\hat{\mathbf{h}}_u^{(t-1)}$ in \eqref{eq:H_est_S}.
    To efficiently estimate $\hat{\mathbf{s}}_u^{(t)}$ by leveraging the near-field sparsity, the virtual channel representation with polar grids as described in Section \ref{subsec:First_stage} are utilized. 
    The grids are dynamically designed in the iterations, where the center of the grids is set as the angle and distance estimates at the previous iteration, and the range of grids decreases with the number of iterations.
    Thus, the angle and distance grids for the $u$-th UE and $l$-th path at the $t$-th iteration are designed as 
    \begin{subequations}
    \begin{align}
        \label{eq:theta_grid_S}
        &\tilde{\theta}_{u,l, g_\theta}^{(t)} \! \in \! \left[ \hat{\theta}_{u,l}^{(t-1)} \!-\! \sigma_\theta^{(t)},\ \hat{\theta}_{u,l}^{(t-1)} \!+\! \sigma_\theta^{(t)}  \right], \\ 
        \label{eq:r_grid_S}
        &\tilde{ r}_{u,l, g_r}^{(t)} \!\! \in  \! \left[ \hat{r}_{u,l}^{(t-1)} \!-\! \sigma_r^{(t)},\ \hat{r}_{u,l}^{(t-1)} \!+\! \sigma_r^{(t)}  \right] ,
    \end{align}
    \end{subequations}
    with $g_\theta \!\in\! \{1, \ldots, \bar{G}_\theta\}$, $g_r \! \in \! \{1, \ldots, \bar{G}_r \}$.
    $\hat{\theta}_{u,l}^{(t-1)}$ and $\hat{r}_{u,l}^{(t-1)}$ are the angle and distance estimates at the $(t-1)$-th iteration, respectively, and $\sigma_\theta^{(t)}$ and $\sigma_r^{(t)}$ are, respectively, the range of angle and distance grids, 
    where the initial values $\hat{\theta}_{u,l}^{(0)}$ and $\hat{r}_{u,l}^{(0)}$ are determined using the angle and distance estimates obtained by the initial channel estimation as shown in Algorithm~\ref{alg:initial}.
    
    Note that the range of angle and distance grids $\sigma_\theta^{(t)}$ and $\sigma_r^{(t)}$ are respectively designed by a monotonically decreasing function such as $\sigma_\theta^{(t)} = a_\theta \exp (-t/2) + b_\theta$ and $\sigma_r^{(t)} = a_r \exp (-t/2) + b_r$ , where the constant values $\{a_\theta, b_\theta, a_r, b_r \}$ are uniquely determined with the desired range $\sigma_\theta^{(1)}$, $\sigma_r^{(1)}$, $\sigma_\theta^{(T)}$, and $\sigma_r^{(T)}$.
    Accordingly, the sets of angle and distance grids for the $u$-th UE are defined as 
    $\tilde{\bm{\theta}}_{u,l}^{(t)} \triangleq \{ \tilde{{\theta}}_{u,l,g_\theta }^{(t)}\}_{g_\theta =1}^{\bar{G}_\theta} $, 
    $\tilde{\bm{\theta}}_{u}^{(t)} \triangleq \{ \tilde{{\bm{\theta}}}_{u,l}^{(t)}\}_{l =1}^{\hat{L}_u} $, 
    $\tilde{\mathbf{r}}_{u,l}^{(t)} \triangleq \{ \tilde{{r}}_{u,l,g_r }^{(t)}\}_{g_r =1}^{\bar{G}_r} $, and
    $\tilde{\mathbf{r}}_{u}^{(t)} \triangleq \{ \tilde{{\mathbf{r}}}_{u,l}^{(t)}\}_{l =1}^{\hat{L}_u} $.

    Using the angle and distance grids in \eqref{eq:theta_grid_S}-\eqref{eq:r_grid_S}, the polar-domain dictionary matrix for the $u$-th UE is designed as 
    \begin{align}
        \label{eq:dictionary_S}
        \tilde{\mathbf{A}}_{u} ( \tilde{\bm{\theta}}_{u}^{(t)}, \tilde{\mathbf{r}}_{u}^{(t)}) = 
        \begin{bmatrix}
            \tilde{\mathbf{A}}_{u,1} ( \tilde{\bm{\theta}}_{u,1}^{(t)}, \tilde{\mathbf{r}}_{u,1}^{(t)}), \ldots,
            \tilde{\mathbf{A}}_{u,\hat{L}_u} ( \tilde{\bm{\theta}}_{u,\hat{L}_u}^{(t)}, \tilde{\mathbf{r}}_{u,\hat{L}_u}^{(t)})
        \end{bmatrix},
    \end{align}
    where 
    $\tilde{\mathbf{A}}_{u,l} ( \tilde{\bm{\theta}}_{u,l}^{(t)}, \tilde{\mathbf{r}}_{u,l}^{(t)}) \in \mathbb{C}^{N \times \bar{G}_\theta \bar{G}_r }$ is the virtual array response for the $u$-th UE and $l$-th path defined as 
    \begin{align*}
        \tilde{\mathbf{A}}_{u,l} ( \tilde{\bm{\theta}}_{u,l}^{(t)}, \tilde{\mathbf{r}}_{u,l}^{(t)}) = 
        \begin{bmatrix}
            \tilde{\mathbf{a}} ( \tilde{{\theta}}_{u,l,1}^{(t)}, \tilde{r}_{u,l,1}^{(t)}), \ldots,
            \tilde{\mathbf{a}} ( \tilde{{\theta}}_{u,l,G_\theta}^{(t)}, \tilde{r}_{u,l,G_r}^{(t)})
        \end{bmatrix}.
    \end{align*}
    
    Through the virtual channel representation with the dictionary matrix $\tilde{\mathbf{A}}_{u} ( \tilde{\bm{\theta}}_{u}^{(t)}, \tilde{\mathbf{r}}_{u}^{(t)})$, the near-field channel for the $u$-th UE can be expressed as 
    \begin{align}
        \label{eq:h_apx_S}
        \mathbf{h}_u 
        = \sum_{l=1}^{\hat{L}} \tilde{\mathbf{A}}_{u,l} (\bm{\theta}_{u,l}^{(t)}, \tilde{\mathbf{r}}_{u,l}^{(t)}) \tilde{\mathbf{z}}_{u,l}
        = \tilde{\mathbf{A}}_{u} ( \tilde{\bm{\theta}}_{u}^{(t)}, \tilde{\mathbf{r}}_{u}^{(t)}) \tilde{\mathbf{z}}_{u},
    \end{align}
    where $\tilde{\mathbf{z}}_{u,l}  \in \mathbb{C}^{\bar{G}_\theta \bar{G}_r \times 1} $ is the virtual path gain vector for the $l$-th path, and 
    $\tilde{\mathbf{z}}_{u} = [ \tilde{\mathbf{z}}_{u,1}^\mathrm{T}, \ldots, \tilde{\mathbf{z}}_{u,\hat{L}_u}^\mathrm{T} ]^\mathrm{T} \in \mathbb{C}^{\bar{G}_\theta \bar{G}_r \hat{L}_u \times 1} $
    is the virtual path gain vector including all paths.

    In light of the near-field model in \eqref{eq:h_apx_S}, an update of the model-based estimate $\hat{\mathbf{s}}^{(t)}_u$ can be obtained by  
    \begin{align}
        \label{eq:s_hat}
        \hat{\mathbf{s}}_u^{(t)} = \mathbf{A} (\hat{\bm{\theta}}_u^{(t)}, \hat{\mathbf{r}}_u^{(t)}) \hat{\mathbf{z}}_{u}^{(t)},
    \end{align}
    with $\hat{\mathbf{z}}_{u}^{(t)}$ denoting the path gain estimates, $\mathbf{A} (\hat{\bm{\theta}}_u^{(t)}, \hat{\mathbf{r}}_u^{(t)})$ being the array responses, which can be computed by solving
    \begin{align}
        \label{eq:min_up_S}
        \underset{\tilde{\mathbf{z}}_u}{\text{minimize}} &
        \ \  \left \| 
            \hat{\mathbf{h}}_u^{(t-1)} - \hat{\mathbf{s}}^{(t)}_u
        \right \|_2^2  \nonumber \\
        \text{subject to} & \ \ \ \hat{\mathbf{s}}^{(t)}_u = \tilde{\mathbf{A}}_{u} ( \tilde{\bm{\theta}}_{u}^{(t)}, \tilde{\mathbf{r}}_{u}^{(t)}) \tilde{\mathbf{z}}_{u} \nonumber \\
        &  \ \  \left \| \tilde{\mathbf{z}}_{u,l} \right \|_{0} = 1, \ \ \forall l \in \{1, 2, \ldots, \hat{L}_u\}.
    \end{align}
    
    Solving the problem \eqref{eq:min_up_S} yields the path gain estimate $\hat{\mathbf{z}}_{u}^{(t)}$, angle estimate $\hat{\bm{\theta}}_u^{(t)}$ and the distance estimate $\hat{\mathbf{r}}_u^{(t)}$, which correspond to the non-zero elements of $\tilde{\mathbf{z}}_u$.

    \subsection{Algorithm Description}

    To illustrate the overall procedure of the proposed JCDE algorithm, 
    the pseudocode is provided in~Algorithm~\ref{alg:EP}, and the algorithmic flow is depicted in Fig.~\ref{fig:alg_flow}.
    As shown in the figure, the algorithm mainly consists of three parts: 1) data estimation, 2) residual error estimation, and 3) reinforcement for the model-based estimate.

    In the data estimation, the mean and variance of the tentative data estimates $\{\hat{x}^q_{u,k}, \xi^{q,x}_{u,k}\}$ are generated through the Soft-IC in~\eqref{eq:mean_IC}-\eqref{eq:cov_IC}, sub-array-wise LMMSE filtering in~\eqref{eq:xq}-\eqref{eq:xiq}, and combining processes~\eqref{eq:qx_prod_mean_vari}, using the received signal $\mathbf{Y}$ and the soft replicas $\{\hat{x}^v_{c,u,k}, \xi^{v,x}_{c,u,k} \}$ and $\{\hat{e}^v_{n_c,u,k}, \xi^{v,e}_{n_c,u,k} \}$.
    Then, the tentative estimates $\{\hat{x}^q_{u,k}, \xi^{q,x}_{u,k}\}$ are updated to $\{\hat{x}_{u,k}, \xi^{x}_{u,k}\}$ through the denoising process in~\eqref{eq:x_denoise}-\eqref{eq:xi_denoise} that exploits the prior $p(x_{u,k})$ in \eqref{eq:prior_x} based on the QAM constellation.
    For the Soft-IC process in the next iteration, the soft replicas $\{\hat{x}^v_{c,u,k}, \xi^{v,x}_{c,u,k} \}$ and $\{\hat{x}^w_{c,u,k}, \xi^{w,x}_{c,u,k} \}$ are updated by suppressing the self-noise feedback in~\eqref{eq:vx_mean_vari} and \eqref{eq:wx_mean}.

    In the residual error estimation, similar to the data estimation, the mean and variance of the tentative residual error estimates $\{ \hat{e}^q_{n_c,u}, \xi^{q,e}_{n_c,u}\}$ are generated through the Soft-IC in~\eqref{eq:mean_IC_e}-\eqref{eq:cov_IC_e} and MRC filtering processes in~\eqref{eq:qe_prod_mean_vari}, using the received signal $\mathbf{Y}$ and the soft replicas $\{\hat{x}^w_{c,u,k}, \xi^{w,x}_{c,u,k} \}$ and $\{\hat{e}^v_{n_c,u,k}, \xi^{v,e}_{n_c,u,k} \}$.
    Then, the tentative estimates $\{ \hat{e}^q_{n_c,u}, \xi^{q,e}_{n_c,u}\}$ are updated to $\{ \hat{e}_{n_c,u}, \xi^{e}_{n_c,u}\}$ through the denoising process in~\eqref{eq:e_denoise} that exploits the prior $p(e_{n_c,u};\mathbf{\Theta})$ in \eqref{eq:prior_e} reflecting the sparsity of the residual errors.
    For the Soft-IC process in the next iteration, the soft replicas $\{\hat{e}_{n_c,u,k}^v, \xi^{v,e}_{n_c,u,k} \}$ are updated by suppressing the self-noise feedback in~\eqref{eq:ve_mean}-\eqref{eq:ve_vari}.
    The hyperparameter $\mathbf{\Theta}$ incorporated in the prior $p(e_{n_c,u}; \mathbf{\Theta})$ is updated through the M-step of the EM algorithm in \eqref{eq:sigma_EM}.
    Based on the updated residual error estimate $\hat{\mathbf{e}}_u$, the channel estimate is reconstructed as $\hat{\mathbf{h}}_u = \hat{\mathbf{s}}_u + \hat{\mathbf{e}}_u$, and the model-based estimate $\hat{\mathbf{s}}_u$ is subsequently refined by solving the optimization problem~\eqref{eq:min_up_S} via OMP~\cite{1993pati_OMP_MIL}.

    After repeating the above procedure for a fixed number of iterations $T$, $\hat{\mathbf{X}}$ and $\hat{\mathbf{H}}$ are adopted as the estimate of the data and channel.
    To enhance convergence performance, a damping scheme \cite{2014Parker_BiGAMP_part1} is introduced in line 6, 7, 12, and 19 of Algorithm~\ref{alg:EP}.
    The damping factor in the proposed method is determined through several simulations by considering the trade-off between estimation performance and convergence rate as in~\cite{2024Ito_BiGaBP, 2019Rangan_VAMP}.
    Note that the proposed initial channel estimation in Algorithm~1 and the JCDE in Algorithm~2 does not require distinguishing between the LoS and NLoS paths.
    
\begin{algorithm}[t]
    \caption[]{Proposed JCDE algorithm}
    \label{alg:EP}
    \hrulefill
    \begin{algorithmic}[1]
        \vspace{-0.5ex}
        \Statex \textbf{Input:} 
            $\mathbf{Y},\ 
            \mathbf{X}_\mathrm{p},\ 
            \hat{\mathbf{H}}^\mathcal{S}_0,\ 
            \{ \hat{\bm{\theta}}_u, \hat{\mathbf{r}}_u \}_{u \in \mathcal{U}},\ 
            \{\hat{L}_u\}_{u=1}^U,\ 
            T,\ \bar{G}_\theta,\ \bar{G}_r,$
            $ \sigma_\theta^{(1)},\ \sigma_\theta^{(T)},\ \sigma_r^{(1)},\ \sigma_r^{(T)}$
        \Statex \textbf{Output:} 
            $\hat{\mathbf{X}},\ \hat{\mathbf{H}},\ \{ \hat{\bm{\theta}}_u,\ \hat{\mathbf{r}}_u \}_{u \in \mathcal{U}} $
        \vspace{-1.5ex}
        \Statex \hspace{-3ex} \hrulefill

        \Statex 
            \textbf{// Initialization} 
        \Statex 
            $\hat{\mathbf{S}} = \mathbf{D}_N \hat{\mathbf{H}}_0^\mathcal{S}$ from Algorithm~\ref{alg:initial} 
        \Statex 
            $\hat{x}^v_{c,u,k_\mathrm{d}} \! = \hat{x}^w_{c,u,k_\mathrm{d}} \! = 0$,\ 
            $\xi^{v,x}_{c,u,k_\mathrm{d}} \! = \xi^{w,x}_{c,u,k_\mathrm{d}} \! = E_s$, $\forall k_\mathrm{d} \in \mathcal{K}_\mathrm{d}$
        \Statex 
            $\hat{x}^v_{c,u,k_\mathrm{p}} \!\!\!\! = \hat{x}^w_{c,u,k_\mathrm{p}} \!\!\!\! = [\mathbf{X}_\mathrm{p}]_{u,k_p}$,\ 
            $\xi^{v,x}_{c,u,k_\mathrm{p}} \!\!\!\! = \xi^{w,x}_{c,u,k_\mathrm{p}} \!\!\!\! = 0$, $\forall k_\mathrm{p} \in \mathcal{K}_\mathrm{p}$
        \Statex 
            $\hat{e}^v_{n_c,u,k} = 0$,\ 
            $\xi^{v,e}_{n_c,u,k} = |[\mathbf{D}_N \hat{\mathbf{H}}_0^\mathcal{S}]_{n_c,u}|^2 $, $\forall k \in \mathcal{K}$

        \For{$t=1, 2, \ldots, T$} 

        \Statex 
            \textbf{\quad // EP for data estimation} 
        \State 
            Calculate $\tilde{\mathbf{y}}^x_{c,u,k},\ \mathbf{\Omega}^{x}_{c,k} $ from \eqref{eq:mean_IC}-\eqref{eq:cov_IC}
        \State 
            Calculate $\hat{x}^{q}_{c,u,k},\ \xi^{q,x}_{c,u,k} $ from \eqref{eq:xq}-\eqref{eq:xiq}
        \State 
            Calculate $\hat{x}_{u,k}^{q},\ \xi_{u,k}^{q,x} $ from \eqref{eq:qx_prod_mean_vari}
        \State 
            Calculate $\hat{x}_{u,k},\ \xi^{x}_{u,k} $ from \eqref{eq:x_denoise}-\eqref{eq:xi_denoise}
        \State 
            Calculate $\hat{x}^{v}_{c,u,k},\ \xi^{v,x}_{c,u,k} $ from \eqref{eq:vx_mean_vari} with damping
        \State 
            Calculate $\hat{x}^{w}_{c,u,k},\ \xi^{w,x}_{c,u,k} $ from \eqref{eq:wx_mean} with damping

        \Statex 
            \textbf{\quad // EP for residual error estimation} 
        \State 
            Calculate $\tilde{y}^e_{n_c,u,k},\ \phi^e_{n_c,u,k}$ from \eqref{eq:mean_IC_e}-\eqref{eq:cov_IC_e}
        \State 
            Calculate $\hat{e}^q_{n_c,u,k},\ \xi^{q,e}_{n_c,u,k}$ from \eqref{eq:qe_mean_vari}
        \State 
            Calculate $\hat{e}_{n_c,u}^{q},\ \xi_{n_c, u}^{q,e} $ from \eqref{eq:qe_prod_mean_vari}
        \State 
            Calculate $\hat{e}_{n_c,u},\ \xi^{e}_{n_c,u} $ from \eqref{eq:e_denoise}
        \State 
            Calculate $\hat{e}^{v}_{n_c,u,k}, \xi^{v,e}_{n_c,u,k}$ from \eqref{eq:ve_mean}-\eqref{eq:ve_vari} with \\ \hspace{3ex} damping
        \State 
            Update channel estimate $\hat{\mathbf{h}}_u = \hat{\mathbf{s}}_u + \hat{\mathbf{e}}_u$ from \eqref{eq:H_est_S}

        \Statex 
            \textbf{\quad // EM algorithm for hyperparameter learning} 
        \State 
            Update $\sigma^e_{n_c,u}$ from \eqref{eq:sigma_EM}

        \Statex 
            \textbf{\quad // Reinforcement for the model-based estimate} 
        \State 
            Generate the grids $\tilde{\theta}_{u,l, g_\theta},\ \tilde{r}_{u,l, g_r} $ from \eqref{eq:theta_grid_S}-\eqref{eq:r_grid_S}
        \State 
            Design the dictionary $\tilde{\mathbf{A}}_{u} ( \tilde{\bm{\theta}}_{u}, \tilde{\mathbf{r}}_{u})$ from \eqref{eq:dictionary_S}
        \State 
            Obtain $\hat{\bm{\theta}}_u, \hat{\mathbf{r}}_u, \hat{\mathbf{z}}_{u}$ by solving \eqref{eq:min_up_S} 
        \State 
            Calculate $\hat{\mathbf{s}}_u$ from \eqref{eq:s_hat} with damping

        \EndFor
            
    \end{algorithmic}
\end{algorithm}

\section{Simulation Results}
    \label{sec:simulation}

    This section evaluates the performance of the proposed initial channel estimation and subsequent JCDE algorithms 
    under the following setup.
    %
    The carrier frequency is $100\ \mathrm{GHz}$,
    the number of BS antennas $N$ is $200$,
    the number of UEs $U$ is $50$, 
    the modulation order $Q$ is $64$-QAM,
    and 
    the length of pilots $K_\mathrm{p}$ and data $K_\mathrm{d}$ are $25$ and $100$, respectively.
    The non-orthogonal pilot 
    $\mathbf{X}_\mathrm{p} \in \mathbb{C}^{50 \times 25}$ 
    is designed by the frame design method in \cite{2022Iimori_bilinear_Grant_free}.
    %
    The near-field channel is composed of $L_u=3$ paths, \textit{i.e.,} $1$ LoS path and $2$ NLoS paths. 
    The total number of paths is $L \!= \! 50 \!\times\! 3 \!=\! 150$, and 
    the corresponding oversampling quantity used in Algorithm~\ref{alg:initial} is set to $\hat{L}=250$.
    The \acp{AoA} and distances are uniformly randomly generated in the range $ \theta_{u,l} \in [-60^\circ, 60^\circ]$ and $r_{u,l} \in [1,10]$~m, respectively. 
    The path gain is generated as
    $z_{u,l} \sim \mathcal{CN}(0, K_\mathrm{f} / (K_\mathrm{f} + 1)),\ (l=1)$ and
    $z_{u,l} \sim \mathcal{CN}(0, 1 / ((K_\mathrm{f} + 1)(L-1))),\ (l \neq 1)$ with a Rician $K$-factor $K_\mathrm{f}=10\ \mathrm{dB}$.
    %
    The polar-domain dictionary $\tilde{\mathbf{A}}( \tilde{\bm{\theta}}, \tilde{\mathbf{r}})$ in \eqref{eq:A_apx} is designed with $G_r = 7$, $G_\theta = 395$ and desired coherence $\bar{\gamma}_\mathrm{coh} = 0.6$ in \cite{2024Xie_URA_2DCoSaMP}.
    %
    The performance is evaluated by the \ac{NMSE} and \ac{BER} under various \ac{SNR}.
    \ac{NMSE} and \ac{SNR} are defined as
    $\mathrm{NMSE}(\bm{\mathbf{H}}) \triangleq \mathbb{E} \left[ {\| \bm{\mathbf{H}} - \hat{\bm{\mathbf{H}}} \|_\mathrm{F}^2} / {\| \bm{\mathbf{H}} \|_\mathrm{F}^2} \right]$, and 
    $\mathrm{SNR} \triangleq
    \mathbb{E} \left[ \| \mathbf{HX} \|_\mathrm{F}^2 \right] / 
    \mathbb{E} \left[ \| \mathbf{N}  \|_\mathrm{F}^2 \right]$.
    In what follows, the initial channel estimation and JCDE performance are evaluated in Section \ref{subsec:sim_initial_CE} and \ref{subsec:sim_JCDE_performance}, respectively.

    \subsection{Initial Channel Estimation Performance}
    \label{subsec:sim_initial_CE}
    %
    To evaluate the initial channel estimation performance, 
    the following estimation methods are compared:
    (a) LS: a classical least squares-based channel estimation,
    (b) \ac{P-SOMP}~\cite{2022Cui_PSOMP}: a near-field channel estimation without considering the non-orthogonality of pilots.  
    (c) 2D-CoSaMP~\cite{2024Xie_URA_2DCoSaMP}: a near-field channel estimation considering non-orthogonality, and 
    (d) the proposed initial channel estimation method in Algorithm~\ref{alg:initial}.

\begin{figure}[t]
        \begin{minipage}{1.0\columnwidth}
            \centering
            \includegraphics[width=\linewidth]{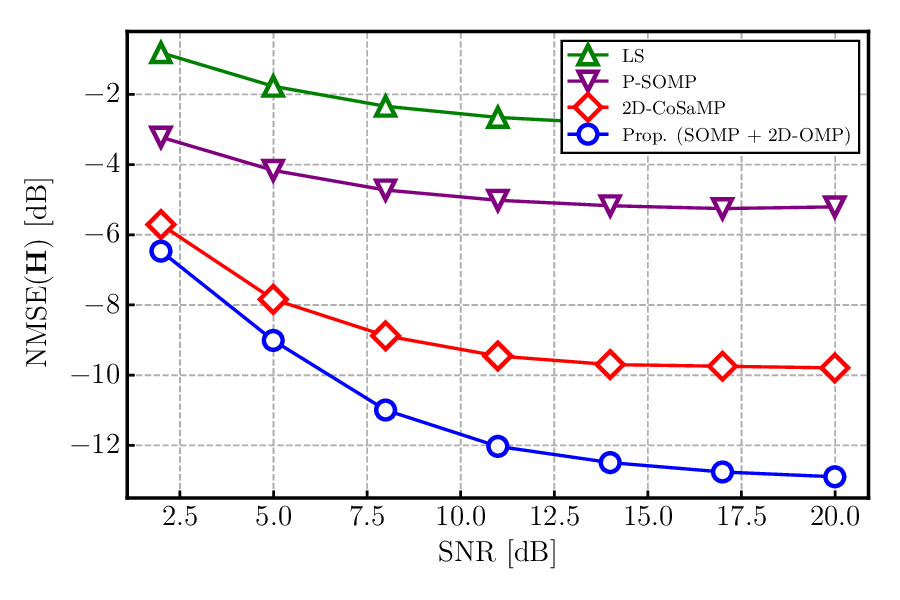}
            \vspace{-0.7cm}
            \subcaption{NMSE versus SNR with the pilot length $K_\mathrm{p}=25$.}
            \label{fig:NMSE_vs_SNR}
        \end{minipage}
        \\
        \begin{minipage}{1.0\columnwidth}
            \vspace{1ex}
            \centering
            \includegraphics[width=\linewidth]{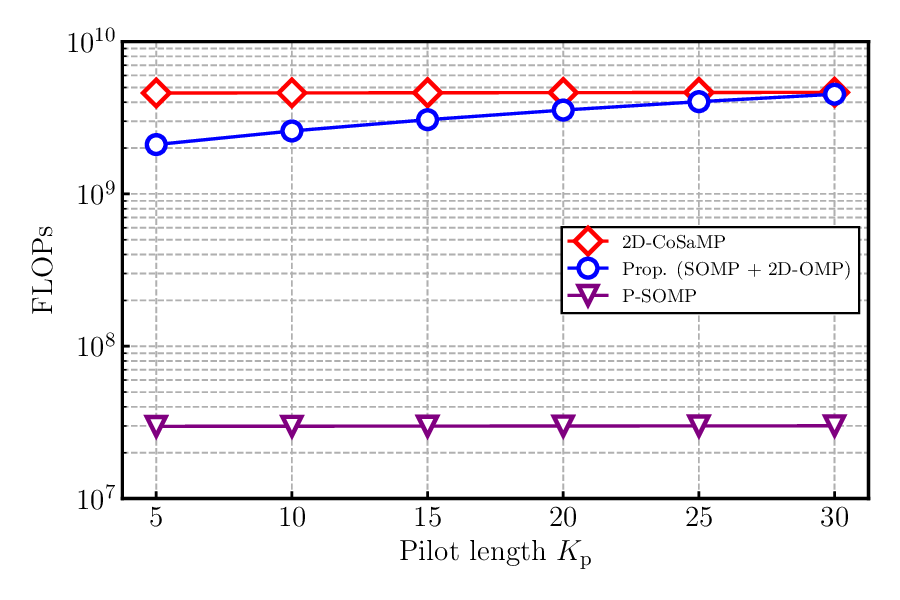}
            \vspace{-0.7cm}
            \subcaption{FLOPs versus the pilot length $K_\mathrm{p}$.}
            \label{fig:FLOPs_vs_Kp}
        \end{minipage}
        \caption{NMSE and FLOPs performance.} 
        \label{fig:Initial}
    \end{figure}

    Fig.~\ref{fig:NMSE_vs_SNR} shows the \ac{NMSE} against \ac{SNR}.
    The \ac{P-SOMP} exhibits limited improvement with an increase in SNR due to pilot contamination stemming from non-orthogonal pilots, whereas 2D-CoSaMP demonstrates a performance enhancement compared to \ac{P-SOMP}.
    The proposed method surpasses these conventional methods by mitigating noise amplification through the utilization of 2D-OMP in the second stage associated with UE-path pairing, resulting in superior channel estimation.
    %
    Fig.~\ref{fig:FLOPs_vs_Kp} and Table~\ref{table:FLOPs_initial} show the computational complexity evaluated by \ac{FLOPs}.
    As depicted in the figure, the \ac{FLOPs} of the proposed method are comparable to 2D-CoSaMP, owing to the two-stage procedure separating angle-distance estimation and UE-path pairing.

    \begin{table}[t!]
        \caption{Computational complexity of initial channel estimation} \label{table:FLOPs_initial}
        \centering
        \begin{tabular*}{8.5cm}{c|c}
            \hline
            \multicolumn{1}{c}{Algotrithm} & FLOPs	\\
            \hline
            \ac{P-SOMP} \cite{2022Cui_PSOMP} & $\mathcal{O}\left(N U(G_r G_\theta + \hat{L}) + \hat{L}^2 U( N + G_r G_\theta )\right)$ 	\\
            2D-CoSaMP \cite{2024Xie_URA_2DCoSaMP} & $\mathcal{O} \left( T_\mathrm{iter} \left( NU(K_\mathrm{p} + G_r G_\theta) + \hat{L} N K_\mathrm{p} + \hat{L}^3 \right ) \right)$ \\
            Proposed & 
            \begin{tabular}{l}
                $\mathcal{O} \Big( N K_\mathrm{p} (G_r G_\theta + \hat{L} + NU)$  \\ 
                $ + \hat{L}^2(G_r G_\theta K_\mathrm{p} + NU + NK_\mathrm{p}) + \hat{L}^3 K_\mathrm{p} + \hat{L}^4 \Big)$ \\
            \end{tabular} \\
            \hline
        \end{tabular*}
        \vspace{1mm}
        \\{Note: $T_\mathrm{iter}$ is the iteration number of 2D-CoSaMP, determined by \cite{2024Xie_URA_2DCoSaMP}}.
    \end{table}

    \subsection{JCDE Performance}
    \label{subsec:sim_JCDE_performance}

    In this subsection, we evaluate the proposed JCDE algorithm.
    %
    As for JCDE algorithm parameters, 
    the damping factor is set to $0.5$, 
    the number of iterations is $T=30$, 
    the number of grids are $\bar{G}_\theta = 5, \bar{G}_r=5$, 
    the grid ranges are 
    $\sigma_\theta^{(1)} = 5^\circ$, 
    $\sigma_\theta^{(T)} = 0.1^\circ$,
    $\sigma_r^{(1)} = 5\ \mathrm{m}$, and  
    $\sigma_r^{(T)} = 1\ \mathrm{m}$, respectively.
    %
    To consider the simplicity of implementation and the complexity of the LMMSE-based filter in \eqref{eq:xq}, the large array with $N$ antennas is evenly divided into $C$ sub-arrays with $N_c=N/C, \ \forall c \in \mathcal{C}$ antennas.
    Since the complexity order of the LMMSE-based filter is $\mathcal{O} \left (K_\mathrm{d} \sum_{c=1}^C N_c^3 + U K_\mathrm{d} \sum_{c=1}^C N_c^2 \right)$, even sub-array partitioning is optimal in terms of minimizing computational complexity given $C$.
    Unless otherwise specified, these parameters are set to $N=200,\ C=4,\ N_c=50$.
    %
    For comparison, AoA-aided BiGaBP \cite{2023Ito_AoA} are employed as a benchmark, which is a state-of-the-art JCDE algorithm.
    Besides, we consider a Genie-aided case with perfect CSI or data corresponding to the lower bound of the proposed method.

    \subsubsection{JCDE Performance with Initial Channel Estimation}
    \label{subsec:sim_JCDE_initial}

    \begin{figure}[t]
        %
        \begin{minipage}{1.0\columnwidth}
            \centering
            \includegraphics[width=\linewidth]{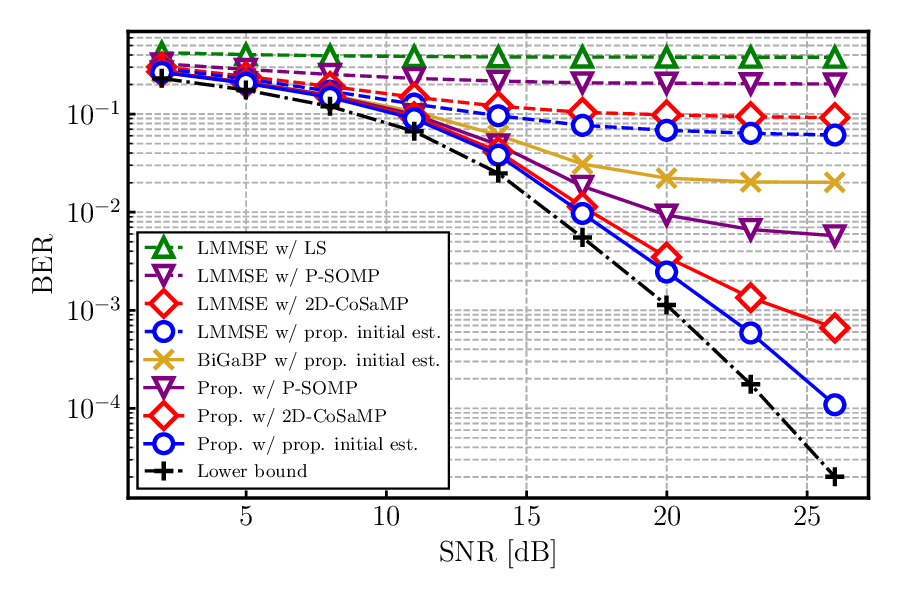}
            \vspace{-0.5cm}
            \subcaption{BER versus SNR with the pilot length $K_\mathrm{p}=25$.}
            \label{fig:BER_vs_SNR_JCDE}
        \end{minipage}
        \\
        \begin{minipage}{1.0\columnwidth}
            \vspace{1ex}
            \centering
            \includegraphics[width=\linewidth]{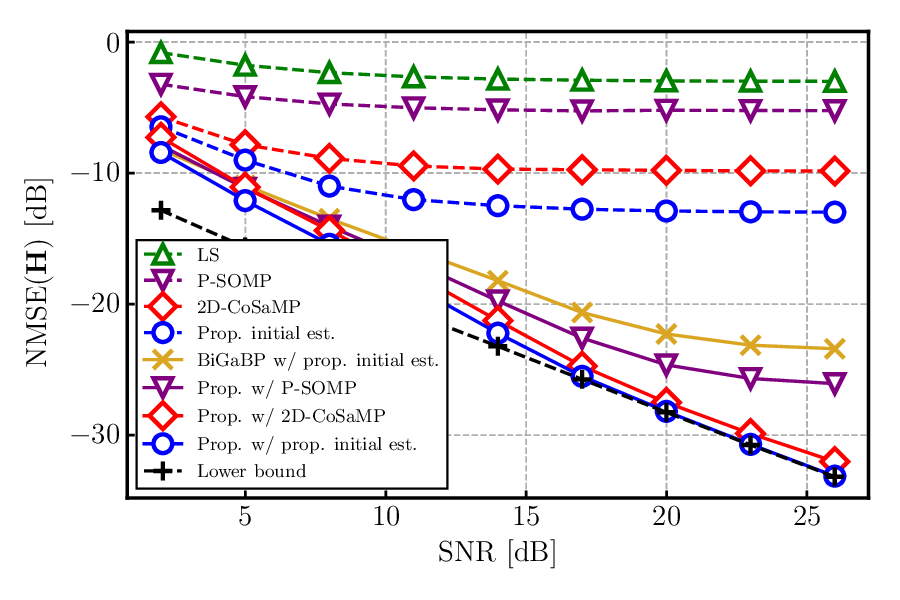}
            \vspace{-0.5cm}
            \subcaption{NMSE versus SNR with the pilot length $K_\mathrm{p}=25$.}
            \label{fig:NMSE_vs_SNR_JCDE}
        \end{minipage}
        \caption{BER and NMSE performance.}
        \label{fig:JCDE_initial}
    \end{figure}
    
    This subsection reveals the NMSE and BER performance of the JCDE algorithms with various initial channel estimation methods, including \ac{P-SOMP}, 2D-CoSaMP, and the proposed initial channel estimation method.
    To evaluate the data detection capability of the above initial channel estimation methods, the LMMSE detector is used for data estimation.
     
    Fig.~\ref{fig:JCDE_initial} shows the BER and NMSE performance.
    As shown in the figures, while LMMSE with LS, which cannot take advantage of the near-field model structures, exhibits poor BER performance, LMMSE with the other initial estimation approaches considering the near-field model structure achieve a slight performance improvement.
    However, there remains high-level error floors due to the non-orthogonal pilots.
    In contrast, the JCDE algorithms boost BER performance due to utilizing both pilot and consecutive data.
    In particular, the proposed JCDE algorithm with the proposed initial channel estimation demonstrates a significant performance gain, approaching the lower bound of perfect \ac{CSI} or perfect data.

    Moreover, the proposed JCDE algorithm demonstrates a notable BER performance compared to the state-of-the-art AoA-aided BiGaBP \cite{2023Ito_AoA}.
    The performance improvement can be attributed to two primary factors.
    The first factor is that the proposed algorithm can leverage the near-field model-based estimation described in Section~\ref{subsec:MB}, whereas \ac{BiGaBP} relies on the far-field assumption.
    The second factor is that the proposed sub-array-wise \ac{LMMSE}-based detection in \eqref{eq:xq} is capable of addressing the correlation between the leaked energy in the beam-domain, whereas BiGaBP is incapable of doing so because of its MRC. 
    To reveal the aforementioned two factors, in Section~\ref{subsec:sim_vs_iterations}, we show the convergence analysis with and without near-field model information.
    Besides, we evaluate in Section~\ref{subsec:sim_vs_C} the performance of the sub-array-wise LMMSE-based detection and its complexity across various numbers of antennas per sub-array~$N_c$.
    
    \subsubsection{Convergence Analysis}
    \label{subsec:sim_vs_iterations}
\begin{figure}[t]
        %
        \begin{minipage}{1.0\columnwidth}
            \centering
            \includegraphics[width=\linewidth]{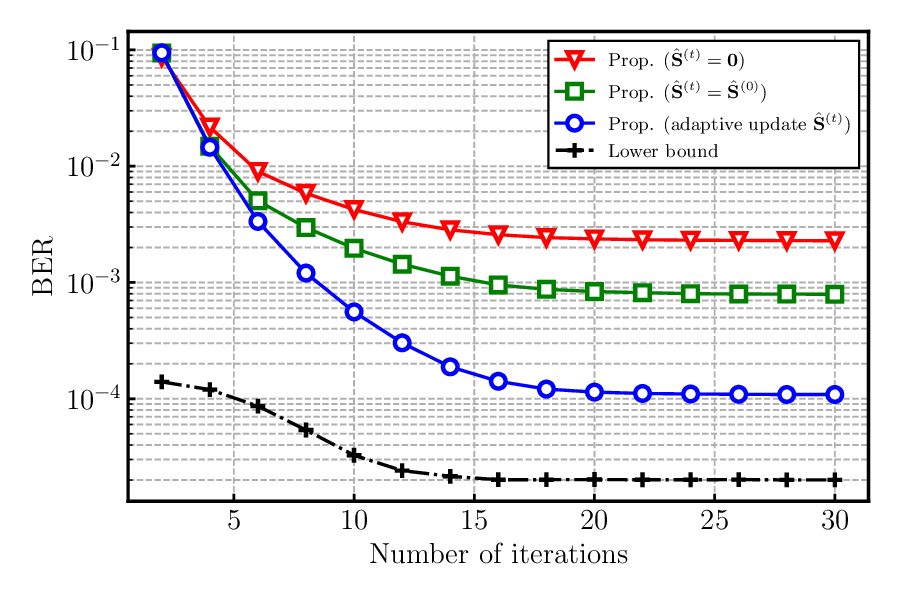}
            \vspace{-0.5cm}
            \subcaption{BER against the number of algorithmic iterations.}
            \label{fig:BER_vs_iterations_JCDE}
        \end{minipage}
        \\
        \begin{minipage}{1.0\columnwidth}
            \vspace{1ex}
            \centering
            \includegraphics[width=\linewidth]{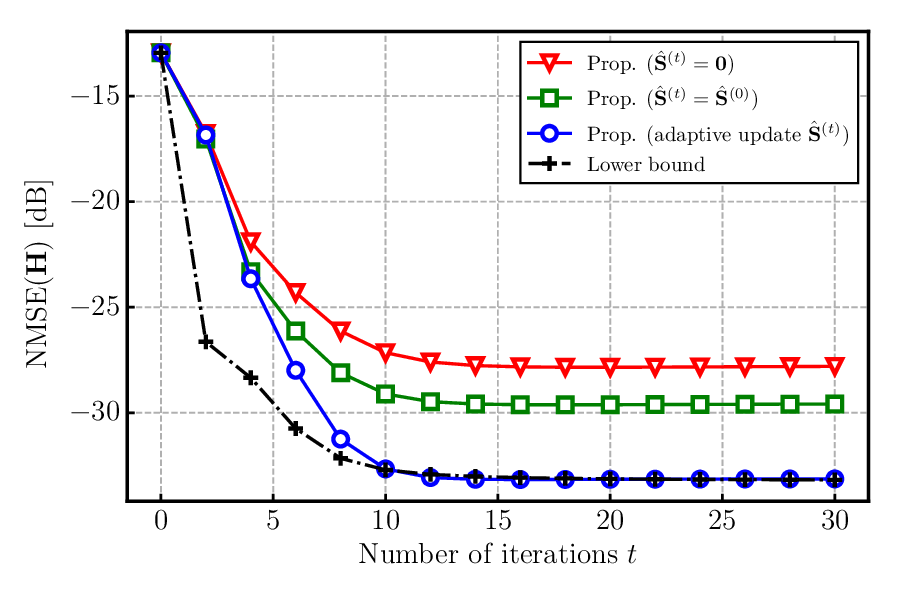}
            \vspace{-0.5cm}
            \subcaption{NMSE against the number of algorithmic iterations.}
            \label{fig:NMSE_vs_iteraions_JCDE}
        \end{minipage}
        \caption{Convergence behavior in term of BER and NMSE performance with $K_\mathrm{p}=25,\ \mathrm{SNR}=26\ \mathrm{dB}$.}
        \label{fig:JCDE_iterarions}
    \end{figure}

    To clarify the advantages gained by leveraging the near-field structure, we evaluate the proposed JCDE algorithm with and without the model-based estimation process explained in Section~\ref{subsec:MB}.
    Fig.~\ref{fig:JCDE_iterarions} illustrates the BER and NMSE convergence behavior with respect to the number of algorithmic iterations.
    In the figure, the red triangle marker corresponds to the proposed JCDE algorithm without the model-based estimate, \textit{i.e.,} $\hat{\mathbf{S}}^{(t)} = \mathbf{0}$, where the prior distribution is designed i.i.d. for each element of $\mathbf{H}$ instead of $\mathbf{E}$, akin to \cite{2019Yan_bilinear,2019Chen_bilinear_cluster}.
    The green square marker corresponds to the proposed JCDE algorithm with the initial model-based estimate but without updating in iterations, \textit{i.e.,} $\hat{\mathbf{S}}^{(t)} = \hat{\mathbf{S}}^{(0)}$.
    Comparing the red triangle maker and green square marker, we can verify the performance improvement owing to the use of the near-filed model through the decomposition of $\mathbf{H}$ into $\hat{\mathbf{S}}$ and $\mathbf{E}$ in \eqref{eq:H_E}.
    Furthermore, in comparison to the proposed algorithm with adaptive update, it can be seen that the adaptive updating of the model-based estimate $\hat{\mathbf{S}}^{(t)}$ enhances the BER and NMSE by exploiting the near-field model.

    \subsubsection{Performance Against the Number of Antennas per Sub-array}
    \label{subsec:sim_vs_C}

    \begin{table}[t!]
        \caption{Computational complexity of JCDE algorithms}  \label{table:FLOPs}
        \centering
        \begin{tabular*}{8.5cm}{l|l}
            \hline
            \multicolumn{1}{l}{Algotrithm} & FLOPs	\\
            \hline
            BiGaBP \cite{2023Ito_AoA} & 
            $\mathcal{O} \Big( U K_\mathrm{d} NQ + UKN + U\hat{L}_u^2N^2 + U N^2 \hat{L}_u \Big)$ 	
            \\
            Proposed & \hspace{-2.4ex}
            \begin{tabular}{l}
                $\mathcal{O}\Big( CUK_\mathrm{d}N_c^2 + C K_\mathrm{d} N_c^3 + U K_\mathrm{d} Q + U K N $
                \\
                \hspace{4ex}$+ U N \bar{G}_\theta \bar{G}_r + U \hat{L}_u^2N + U\hat{L}_u \bar{G}_\theta \bar{G}_r \Big)$ 	
            \end{tabular} 
            \\
            \hline
        \end{tabular*}
    \end{table}

    \begin{figure}[t]
        %
        \begin{minipage}{1.0\columnwidth}
            \centering
            \includegraphics[width=\linewidth]{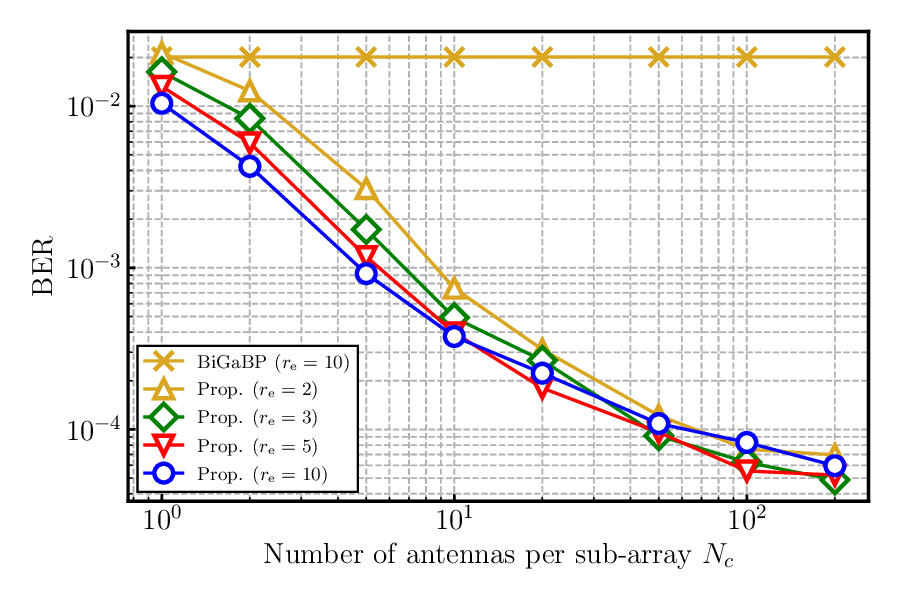}
            \vspace{-0.5cm}
            \subcaption{BER against the number of antennas per sub-array $N_c$.
            To investigate the impact of energy leakage effects, the distances between the BS and the $u$-th UE are uniformly randomly generated within the range $r_{u,l} \in [1, r_\mathrm{e}]$ m.}
            \label{fig:BER_vs_C_JCDE}
        \end{minipage}
        \\
        \begin{minipage}{1.0\columnwidth}
            \vspace{1ex}
            \centering
            \includegraphics[width=\linewidth]{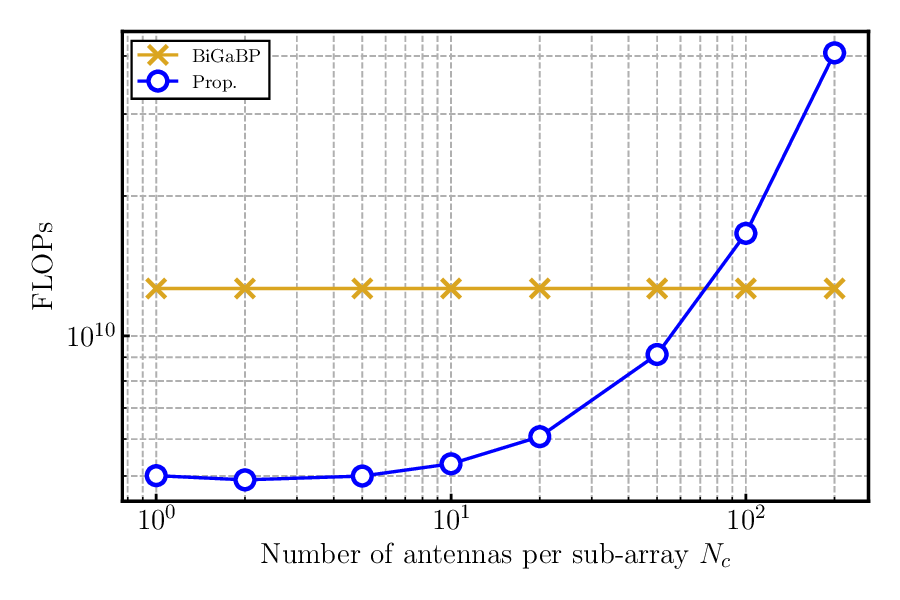}
            \vspace{-0.5cm}
            \subcaption{FLOPs against the number of antennas per sub-array $N_c$ with $N=200$.}
            \label{fig:FLOPs_vs_C_JCDE}
        \end{minipage}
        \\
        \begin{minipage}{1.0\columnwidth}
            \vspace{1ex}
            \centering
            \includegraphics[width=\linewidth]{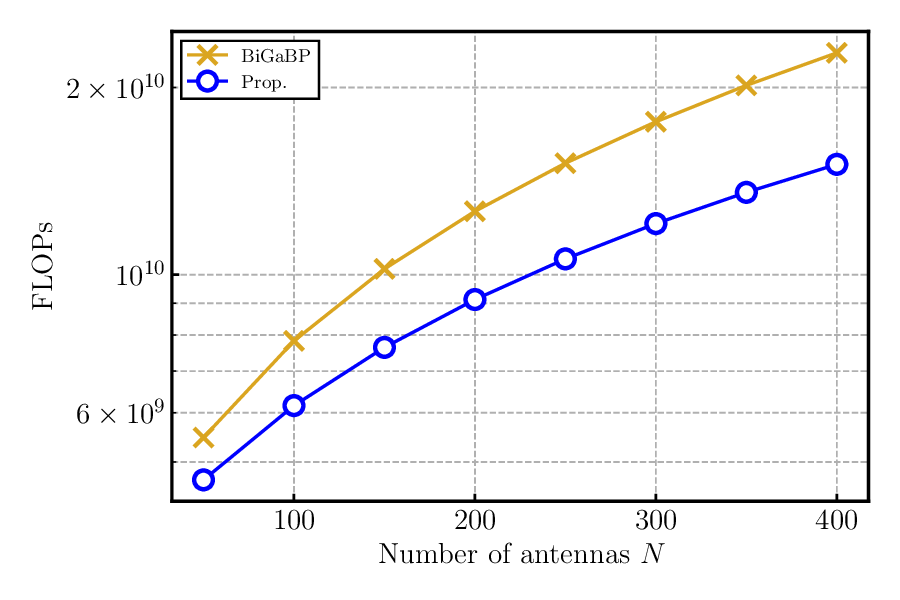}
            \vspace{-0.5cm}
            \subcaption{FLOPs against the number of antennas $N$ with $N_c=50$.}
            \label{fig:FLOPs_vs_N_JCDE}
        \end{minipage}
        \caption{BER and FLOPs performance with $K_\mathrm{p}=25,\ \mathrm{SNR}=26\ \mathrm{dB}$.}
        \label{fig:performance_vs_C}
    \end{figure}

    To analyze the impact of the number of antennas per sub-arrays $N_c$ on the performance of the JCDE algorithm employing the sub-array-wise LMMSE-based detection, we offer in Fig.~\ref{fig:performance_vs_C} the BER and FLOPs with respect to various numbers of antennas per sub-array $N_c$, where $N_c=N=200$ corresponds to the full-array LMMSE-based detection and $N_c=1$ corresponds to the MRC-based detection.
    As depicted in the figure, the BER decreases as the number of antennas per sub-array $N_c$ increases 
    (\textit{i.e.,} the number of sub-arrays $C$ decreases) 
    because each sub-array effectively address the correlation in the beam-domain caused by energy leakage as $N_c$ increases.
    In particular, the MRC-based detection corresponding to $N_c=1$ exhibits  poor performance.
    %
    To analyze the energy leakage effects under various intensities of near-field effects, the BS-UE distance $r_\mathrm{e}$ is varied across different values, where the distances between the BS and the $u$-th UE are uniformly randomly generated within the range $r_{u,l} \in [1, r_\mathrm{e}]\ \mathrm{[m]}$.
    As shown in the figure, when $N_c$ is small, the BER performance deteriorates as the BS-UE distance decreases. This is because each sub-array with a small number of antennas cannot capture the leaked energy that grows more prominent at shorter distances, as illustrated in Fig~\ref{fig:NF_DFT}.
    %
    In contrast, a decrease in the number of antennas per sub-array leads to a reduction in FLOPs attributed to the decreased size of the inverse matrix associated with the LMMSE-based detection in \eqref{eq:xq}.
    Despite relying on the LMMSE-based detector, the proposed algorithm can achieve lower FLOPs when $N_c \leq 50\ (C \geq 4)$ compared to BiGaBP, which relies on an MRC-based detector, since the proposed method suppresses self-feedback in \eqref{eq:vx_mean_vari} after the denoising process as in \eqref{eq:x_denoise}-\eqref{eq:xi_denoise} with FLOPs $\mathcal{O}(U K_d Q)$, whereas BiGaBP suppresses self-feedback before the denoising process \cite{2023Ito_AoA,2022Tamaki_GAMP_feedback} with FLOPs $\mathcal{O}(N U K_d Q)$ that is dominant complexity throughout the entire process as shown in Table~\ref{table:FLOPs}.
    As a result, the proposed method achieves lower computational complexity across various numbers of antennas $N$ compared to the conventional method, as shown in Fig.~\ref{fig:FLOPs_vs_N_JCDE}.
    From the above results, it is evident that the proposed method outperforms the conventional method
    in both data detection and complexity.

\section{Conclusion}
    This paper proposed an initial channel estimation algorithm and subsequent JCDE algorithm for multiuser XL-MIMO systems with non-orthogonal pilots.
    The initial channel estimation is performed by an efficient two-stage compressed sensing algorithm exploiting the polar-domain sparsity.
    Furthermore, the initial channel estimates are refined by jointly utilizing both non-orthogonal pilots and data via the EP algorithm.
    To improve channel estimation accuracy, the model-based deterministic approach is integrated into a Bayesian inference framework.
    In addition, to address the near-field specific correlation in the beam-domain, a sub-array-wise LMMSE filter is designed considering the correlation and channel estimation errors for data detection.
    Computer simulations validated that the proposed method is superior to existing approaches in terms of channel estimation, data detection, and complexity.
     
    For future work, a flexible sub-array partitioning scheme, where each sub-array has a different number of antennas, should be explored to more effectively mitigate the leaked energy in the beam-domain.

\bibliographystyle{IEEEtran}
\bibliography{reference.bib}

\end{document}